\documentclass[preprint,12pt]{elsarticle}

\usepackage{graphicx}
\usepackage{subfigure}
\usepackage{amsmath,amssymb,bm}
\usepackage{gensymb}

\usepackage{lineno}

\usepackage{multirow}
\usepackage{color,soul}

\usepackage{algorithm,algcompatible,amsmath}
\algnewcommand\INPUT{\item[\textbf{Input:}]}%
\algnewcommand\OUTPUT{\item[\textbf{Output:}]}%

\usepackage{nomencl}
\makenomenclature

\usepackage{etoolbox}
\renewcommand\nomgroup[1]{%
  \item[
  \ifstrequal{#1}{A}{\textit{Abbreviations}}{%
  \ifstrequal{#1}{S}{\textit{Sets and indices}}{%
  \ifstrequal{#1}{P}{\textit{Given parameters}}{%
  \ifstrequal{#1}{E}{\textit{Estimated parameters}}{%
  \ifstrequal{#1}{F}{\textit{Functions}}{%
  \ifstrequal{#1}{V}{\textit{Variables}}{}}}}}}%
]}

\journal{Applied Energy}

\begin{document}

\begin{frontmatter}


\title{Optimization-based modelling and game-theoretic framework for techno-economic analysis of demand-side flexibility: a real case study}



\author{Timur Sayfutdinov\fnref{label1}\corref{cor1}}
\author{Charalampos~Patsios\fnref{label1}}
\author{David~Greenwood\fnref{label1}}
\author{Meltem~Peker\fnref{label1}}
\author{Ilias~Sarantakos\fnref{label1}}

\address[label1]{School of Engineering, Newcastle University, Newcastle upon Tyne, United Kingdom}
\cortext[cor1]{t.saifutdinov@newcastle.ac.uk}

\begin{abstract}
This paper proposes a two-step framework for techno-economic analysis of a demand-side flexibility service in distribution networks. Step one applies optimization-based modelling to propose a generic problem formulation which determines the offer curve, in terms of available flexible capacity and its marginal cost, for flexible distribution-connected assets. These offer curves form an input to the second step, which uses a multi-agent iterative game framework to determine the benefits of demand-side flexibility for the Distribution System Operator (DSO) and the service providers. The combined two-step framework simultaneously accounts for the  objectives of each flexibility provider, technical constraints of flexible assets, customer preferences, market clearing mechanisms, and strategic bidding by service providers, omission of any of which can lead to erroneous results. The proposed two-step framework has been applied to a real case study in the North East of England to examine four market mechanisms and three bidding strategies. The results showed that among all considered market mechanisms, flexibility markets that operate under discriminatory pricing, such as pay-as-bid and Dutch reverse auctions, are prone to manipulations, especially in the lack of competition. In contrast, uniform pricing pay-as-cleared auction provides limited opportunities for manipulation even when competition is low. 
\end{abstract}


\end{frontmatter}


\mbox{}
\nomenclature[A]{GB}{Great Britain}
\nomenclature[A]{DSO}{distribution system operator}
\nomenclature[A]{RES}{renewable energy source}
\nomenclature[A]{EES}{electrical  energy  storage}
\nomenclature[A]{DER}{distributed energy resources}
\nomenclature[A]{LCT}{low carbon technology}
\nomenclature[A]{HP}{heat pump}
\nomenclature[A]{EV}{electric vehicle}
\nomenclature[A]{DNO}{distribution network operator}
\nomenclature[A]{DUoS}{distribution use of system}
\nomenclature[A]{ToU}{time of use}
\nomenclature[A]{VCG}{Vickrey-Clarke-Groves}
\nomenclature[A]{ESO}{electricity system operator}
\nomenclature[A]{MA}{multi-agent}
\nomenclature[A]{VPP}{virtual power plant}
\nomenclature[A]{I\&C}{industrial and commercial}
\nomenclature[A]{DSR}{demand-side response}
\nomenclature[A]{PAB}{pay-as-bid}
\nomenclature[A]{PAC}{pay-as-cleared}
\nomenclature[A]{DRA}{Dutch reverse auction}
\nomenclature[A]{DoD}{depth of discharge}
\nomenclature[A]{MILP}{mixed-integer linear programming}
\nomenclature[A]{MCP}{market clearing price}
\nomenclature[A]{OP}{overpricing}
\nomenclature[A]{US}{understatement}
\nomenclature[A]{UB}{underbidding}
\nomenclature[A]{ST}{system transformation}
\nomenclature[A]{CT}{customer transformation}
\nomenclature[A]{LW}{leading the way}
\nomenclature[A]{NZE}{net-zero early}
\nomenclature[A]{Det.}{detached}
\nomenclature[A]{S-Det.}{semi-detached}
\nomenclature[A]{Ter.}{terraced}
\nomenclature[A]{JuMP}{Julia for mathematical programming}
\nomenclature[A]{CP}{convex programming}
\nomenclature[A]{LP}{linear programming}
\nomenclature[A]{ENA}{energy networks association}
\nomenclature[A]{CLDS}{customer-led distribution system}

\nomenclature[S]{\(M\)}{set of constraints, indexed by $m$}
\nomenclature[S]{\(T\)}{set of time intervals (modelling time horizon), indexed by $t$}
\nomenclature[S]{\(T^\text{FW} \in T\)}{flexibility time window (subset of time intervals $T$), indexed by $t$}
\nomenclature[S]{\(D\)}{set of dwelling types, indexed by $d$}
\nomenclature[S]{\(I\)}{set of linear segments, indexed by $i$}
\nomenclature[S]{\(T^\text{Rec} \in T\)}{load recovery period (subset of time intervals $T$), indexed by $t$}
\nomenclature[S]{\(A\)}{set of agents, indexed by $a$}
\nomenclature[S]{\(G\)}{set of price levels, indexed by $g$}
\nomenclature[S]{\(N\)}{set of demand-side flexibility types, indexed by $n$}

\nomenclature[P]{\(\epsilon\)}{tolerance value}
\nomenclature[P]{\(\pi\)}{availability fee, \textsterling/MW/hour}
\nomenclature[P]{\(\Delta \text{T}^\text{FW}\)}{duration of flexibility time window, hours}
\nomenclature[P]{\(\text{P}^\text{Org}_t\)}{original demand consumption of a flexible asset, MW}
\nomenclature[P]{\(\text{N}^\text{HP}\)}{number of households equipped with HPs, units}
\nomenclature[P]{\(\mu_d\)}{share of a dwelling type, pu}
\nomenclature[P]{\(\text{c}^\text{En}_t\)}{Electricity tariff (e.g., ToU), \textsterling/MWh}
\nomenclature[P]{\(\Delta t\)}{time resolution (time step), hours}
\nomenclature[P]{\(\text{c}^\text{Pen}\)}{specific value of customer discomfort, \textsterling/\degree C$^2$/hour or \textsterling/MWh$^2$/hour}
\nomenclature[P]{\(\bar{\text{P}}_d^\text{HP}\)}{HP average power rating, MW}
\nomenclature[P]{\(\text{K}^\text{Th}_d\)}{effective thermal conductance of a dwelling, MW/\degree C}
\nomenclature[P]{\(\text{C}^\text{Th}_d\)}{effective thermal  capacitance  of a dwelling, MWh/\degree C}
\nomenclature[P]{\(\tau^\text{Amb}_t\)}{ambient  temperature  profile, \degree C}
\nomenclature[P]{\(\text{k}^\text{HP}\)}{HP conversion factor, pu}
\nomenclature[P]{\(\text{p}^\text{HP}\)}{heating peak demand factor (ratio of peak demand for heating to daily average consumption for heating), pu}
\nomenclature[P]{\(\Delta \text{T}\)}{duration of modelling time horizon, hours}
\nomenclature[P]{\(\tau^\text{Min},\tau^\text{Max}\)}{minimum and maximum comfortable temperature values, \degree C}
\nomenclature[P]{\(\text{N}^\text{Plg}_t\)}{number of plugged in EVs, units}
\nomenclature[P]{\(\text{N}^\text{EV}\)}{total number of EVs, units}
\nomenclature[P]{\(\bar{\text{P}}^\text{EV}\)}{aggregate capacity of EV chargers, MW}
\nomenclature[P]{\(\text{R}^\text{In}\)}{battery internal resistance, Ohm}
\nomenclature[P]{\(\text{V}^\text{OC}\)}{battery  open-circuit  voltage, V}
\nomenclature[P]{\(\text{e}^\text{EV}\)}{total daily energy demand from the aggregation of EVs, MWh}
\nomenclature[P]{\(\text{P}^\text{EVDem}_t\)}{original EV charging demand profile, MW}
\nomenclature[P]{\(\bar{\text{E}}\)}{EES installed energy capacity, MWh}
\nomenclature[P]{\(\text{C}^\text{E}\)}{price for the installed energy capacity of EES, \textsterling/MWh}
\nomenclature[P]{\(\text{N}^\text{Cyc}_i\)}{EES cycle lifetime, cycles}
\nomenclature[P]{\(\bar{\text{P}}^\text{ES}\)}{EES installed power capacity, MW}
\nomenclature[P]{\(\eta^\text{Ch},\eta^\text{Dis}\)}{EES charge and discharge efficiencies, pu}
\nomenclature[P]{\(\underline{\text{DoD}}_i, \overline{\text{DoD}}_i\)}{low and high bounds of the linear segment $i$, pu}
\nomenclature[P]{\(\text{e}^\text{I\&C}\)}{energy recovery factor, pu}
\nomenclature[P]{\(\text{p}^\text{I\&C}\)}{power recovery factor, pu}
\nomenclature[P]{\(\bar{\text{P}}^\text{I\&C}\)}{I\&C DSR capacity, MW}
\nomenclature[P]{\(\bar{\pi}\)}{ceiling price, \textsterling/MW/hour}
\nomenclature[P]{\(\text{P}^\text{D}\)}{flexible capacity demand, MW}
\nomenclature[P]{\(\textbf{P}^\textbf{True}_a\)}{vector of agent's true offers, MW}
\nomenclature[P]{\(\bm{\pi}^\textbf{True}_a\)}{vector of agent's marginal costs, \textsterling/MW/hour}
\nomenclature[P]{\(\bm{E}\)}{vector of ones, pu}
\nomenclature[P]{\(\pi^\text{DM}_g\)}{value of price level $g$, \textsterling/MW/hour}
\nomenclature[P]{\(\pi^\text{Start}_a\)}{agent's starting price, \textsterling/MW/hour}
\nomenclature[P]{\(\text{a}^\text{I\&C},\text{b}^\text{I\&C}\)}{quadratic and linear terms' coefficients of I\&C cost function, \textsterling/MW$^2$, \textsterling/MW}
\nomenclature[P]{\(\text{N}^\text{A}\)}{total number of agents, units}

\nomenclature[E]{\(\bm{\Delta \pi}\)}{vector of changes in offered prices, \textsterling/MW/hour}
\nomenclature[E]{\(\bm{\Delta P}\)}{vector of changes in offered capacities, MW}
\nomenclature[E]{\(\bm{\pi}^\textbf{O}\)}{vector of offer prices, \textsterling/MW/hour}
\nomenclature[E]{\(\textbf{P}^\textbf{O}\)}{vector of submitted offers, MW}
\nomenclature[E]{\(\Pi_a\)}{agent's calculated profit, \textsterling}
\nomenclature[E]{\(\Delta \pi_a\)}{agent's price update, \textsterling/MW/hour}
\nomenclature[E]{\(\Delta P_a\)}{agent's offered capacity update, MW}
\nomenclature[E]{\(\text{P}^\text{DM}_g\)}{flexibility demand met at price level $g$, MW}
\nomenclature[E]{\(\Pi^\text{UB}_a\)}{expected profit from underbidding, \textsterling}
\nomenclature[E]{\(\sigma\)}{share of flexible capacity, pu}

\nomenclature[F]{\(h_m(\cdot)\)}{technical constraints of flexible assets}
\nomenclature[F]{\(\lvert \cdot \rvert\)}{cardinality of a set}
\nomenclature[F]{\(C^\text{I\&C}(\cdot)\)}{cost function for I\&C DSR, \textsterling}

\nomenclature[V]{\(P^\text{F} \in \mathbb{R}_{\geq 0}\)}{flexible capacity, MW}
\nomenclature[V]{\(P^\text{Sch}_t \in \mathbb{R}\)}{scheduled power consumption/generation of a flexible asset, MW}
\nomenclature[V]{\(\bm{x} \in \mathbb{R}^n\)}{array of decision variables}
\nomenclature[V]{\(P^\text{HP}_{d,t} \in \mathbb{R}_{\geq 0}\)}{heating demand, MW}
\nomenclature[V]{\(\tau^\text{P}_{d,t},\tau^\text{N}_{d,t} \in \mathbb{R}_{\geq 0}\)}{positive and negative deviations from the comfortable indoor temperature range, \degree C}
\nomenclature[V]{\(\tau_{d,t} \in \mathbb{R}_{\geq 0}\)}{dwelling indoor temperature, \degree C}
\nomenclature[V]{\(E^\text{EV}_t \in \mathbb{R}_{\geq 0}\)}{aggregate state of energy of EVs, MWh}
\nomenclature[V]{\(P^\text{EVT}_t \in \mathbb{R}_{\geq 0}\)}{charging power at terminals of EVs, MW}
\nomenclature[V]{\(P^\text{EVB}_t \in \mathbb{R}_{\geq 0}\)}{charging power at batteries of EVs, MW}
\nomenclature[V]{\(E^\text{EVL}_t \in \mathbb{R}_{\geq 0}\)}{EV charging demand not met, MWh}
\nomenclature[V]{\(E^\text{ES}_t \in \mathbb{R}_{\geq 0}\)}{EES state of energy, MWh}
\nomenclature[V]{\(P^\text{Ch}_t \in \mathbb{R}_{\geq 0}\)}{EES charging power, MW}
\nomenclature[V]{\(P^\text{Dis}_t \in \mathbb{R}_{\leq 0}\)}{EES discharging power, MW}
\nomenclature[V]{\(DoD \in \mathbb{R}_{[0,1]}\)}{EES depth of discharge, pu}
\nomenclature[V]{\(\alpha_i \in \mathbb{Z}_{[0,1]}\)}{auxiliary binary variable}
\nomenclature[V]{\(P^\text{I\&C}_t\)}{I\&C flexible power consumption, MW}
\nomenclature[V]{\(\bm{P}^\textbf{A} = \{\bm{P}^\textbf{A}_a \in \mathbb{R}^{\lvert G \rvert}_{\geq 0}\}\)}{vector of accepted offers, MW}
\nomenclature[V]{\(x \in \mathbb{R}_{[0,\text{P}^\text{D}]}\)}{auxiliary variable (unmet flexibility demand, MW)}
\nomenclature[V]{\(\lambda \in \mathbb{R}_{\geq 0}\)}{market clearing price (dual variable), \textsterling/MW/hour}
\textcolor{black}{
\printnomenclature}

\section{Introduction}\label{sec:introduction}

\subsection{Motivation}
Electrification of domestic heating, transport, and industry, coupled with the increasing penetrations of Renewable Energy Sources (RESs) will require substantial transformations of the existing electricity distribution infrastructure. Many options are available to respond to this challenge, including network reinforcement and wide adoption of Electrical Energy Storage (EES); however, demand-side flexibility is expected to be one of the most cost-effective solutions \cite{strbac2020role}. 
Increasing levels of Distributed Energy Resources (DERs) in distribution networks and dispatchable Low Carbon Technologies (LCTs), such as Heat Pumps (HPs), Electric Vehicles (EVs), and EES, will allow future networks to operate in a more efficient and sustainable way. Dispatching DERs and LCTs at the distribution level can not only defer reinforcement of the local network, but also contribute to achieving system wide goals, such as net zero emissions \cite{bouckaert2021net}. Acknowledging this, most European countries -- which have set ambitious emission reduction targets -- have introduced the concept of the Distribution System Operator (DSO). DSO refers to endowing Distribution Network Operators (DNOs) with additional responsibilities, requiring them to take more active control over their networks \cite{ena2018future,eudso2021}. In addition to traditional network flexibility assets (e.g., on-load tap changing transformers, EES, and capacitor banks), DSOs are encouraged to facilitate various types of demand-side flexibility, including price-driven flexibility (e.g., by Distribution Use of System (DUoS) charges, Time of Use (ToU) and dynamic tariffs), contracted flexibility from electricity customers, as well as local energy trading \cite{en2017open}. 

Demand-side flexibility, which refers to the ability of demand-side assets to change their consumption to meet system needs, is a topic of great interest to researchers, DSOs, regulators, aggregators, and potential flexibility providers. In the existing literature, some researchers develop methods to integrate demand-side flexibility services into distribution networks, whilst others model flexible assets (i.e., EES, HP, and EV) and propose various \textcolor{black}{participation} algorithms for flexibility service providers. \textcolor{black}{However, a comprehensive method to evaluate the costs and benefits of the DSO procured flexibility service to stakeholders within the local network, which accounts for rational objective-driven behavior (i.e., bidding) of individual service providers, has not been proposed yet.} In this paper, we have developed a comprehensive framework for the techno-economic analysis of contracted demand-side flexibility service, which accounts for particular market mechanisms, market manipulation by flexibility providers, technical constraints of flexible assets, customer preferences, and costs associated with the service delivery, including ToU tariffs. The necessity of considering the factors above is dictated by their significant influence to the problem inputs and outputs. 
\textcolor{black}{For instance, we explicitly show in the results that depending on the particular market clearing mechanism and number of service providers, the market clearing price for the service can vary by as much as 87.5\% (for the cases considered) due to the bidding, leading to the corresponding variation of DSO calculated costs. The latter illustrates the significance of modelling DSO flexibility services cognizant of providers' bidding strategies. In addition, }
the customer preferences and technical constraints of flexible assets significantly affect the amount of flexible capacity that can be provided from these devices. For example, an increase in ambient temperature of 1\degree C may lead to more than 5\% reduction in the flexible capacity provided by HPs. Finally, ToU tariffs directly affect the costs incurred by service providers; this in turn affects their offer curve, and therefore the price of the service.

The proposed framework can be used to evaluate the technical potential and economic effect of a prospective demand-side flexibility service and to inform the selection of an appropriate market mechanism.  

\subsection{Literature review}
Two relevant streams can be highlighted in the literature on demand-side flexibility related to the present study. 
The first considers the design of flexibility markets to integrate flexibility into distribution networks, while the second focuses on modelling of flexible assets and proposes bidding strategies for flexibility providers.

In the literature on flexibility market design, the authors propose various schemes for enabling contracted flexibility markets. One approach, proposed by Samadi \textit{et al.} \cite{samadi2012advanced}, uses the Vickrey-Clarke-Groves (VCG) auction for demand-side management in smart grids. In the VCG auction, each participant is remunerated according to the positive effect it causes to the system, which eliminates any possibility for market manipulations. The VCG mechanism is also used by Seklos \textit{et al.} \cite{seklos2020designing} to calculate pay-off for flexibility providers in a reactive power flexibility market. Even though the VCG mechanism is resistant to market manipulations, it has not been widely adopted in practice due to numerous issues, which are explained by Rothkopf \textit{et al.} in \cite{rothkopf2007thirteen}. A framework for DSOs to aggregate and provide flexibility to Electricity System Operator (ESO) is proposed by Chen \textit{et al.} \cite{chen2019aggregate}. The authors provide models to quantify the aggregate power flexibility from DERs and apply these to a model predictive control approach. In \cite{tsaousoglou2021mechanism}, Tsaousoglou \textit{et al.} propose a DSO flexibility market framework to maintain network voltages and currents within operational limits. In this paper the authors look at the fairness of pay-off to flexibility providers based on their level of effect on the community's electricity cost, which is similar to the VCG mechanism. An iterative auction scheme for a community of prosumers to resolve network constraints is proposed by Tsaousoglou \textit{et al.} \cite{tsaousoglou2021transactive}. The auction algorithm is developed to align the objectives of individual customers with the global objective, making it strategy-resistive. In \cite{kanellos2017power}, Kanellos \textit{et al.} propose a port power management method using a Multi-Agent (MA) system design. In the proposed framework, the global objective is achieved by aligning the individual objectives of agents, which is a valid assumption when the system is managed by a single entity. However, in the case of the independent (greedy) agents, such objectives' alignment becomes a non-trivial problem. In \cite{wang2019multi}, Wang \textit{et al.} apply an MA approach to incorporate residential demand-side flexibility into the electricity market. In the proposed framework, agents follow the solutions of specific optimization problems to achieve individual objectives which do not necessarily align with the global objective. 

In the literature on bidding strategies for flexibility providers, the research questions focus on methods and algorithms which allow flexibility providers to maximise their own profits. 
In \cite{rahimiyan2015strategic}, Rahimiyan \textit{et al.} propose a strategic bidding approach for a Virtual Power Plant (VPP) participating in the day-ahead and real-time energy markets. In the proposed framework, the VPP manages a cluster of price-responsive demands, wind power plants, and EES. Its strategic decision-making is based on a two-stage (day-ahead and real-time) robust optimization procedure. 
An MA system approach has been used by Reis \textit{et al.} in  \cite{reis2020multi} to model an energy community in which demand-side flexibility was applied to maximize the community sustainability by implementing detailed modelling of customers and following an optimization heuristic for agents' decision-making.
Golmohamadi \textit{et al.} have applied an MA system modelling approach to study the behaviour of residential and industrial customers participating in market-based demand-side flexibility services \cite{golmohamadi2019multi}. The approach is based on the individual objectives of agents (aggregators, industrial and domestic customers), which are determined by means of formal optimization problems.
A risk-constrained bidding strategy for demand-side flexibility has been proposed in \cite{afzali2020risk}, where Afzali \textit{et al.} considered the participation of DER and EV aggregators in day-ahead electricity market. 
In \cite{wang2020aggregation}, Wang \textit{et al.} proposed an optimization-based bidding method for demand-side flexibility aggregators in electricity markets. The authors have illustrated how complete information about the environment can be used by strategic aggregators to manipulate the market, reducing system efficiency while maximising their own benefits. Finally, a bidding strategy for a simple two aggregator system, based on robust optimization and elements of game theory, has been proposed by Abapour \textit{et al.} in \cite{abapour2020robust}.


\subsection{Research gaps}

The literature reviewed on flexibility market design shows a lack of studies on the conventional market mechanisms (e.g., pay-as-bid, pay-as-cleared) which have proven their efficacy for existing electricity markets and services. \textcolor{black}{For instance, in the Great Britain (GB), the wholesale electricity market is cleared via pay-as-cleared, the capacity market is cleared via reverse auction, the balancing mechanism and some ESO ancillary services are cleared via pay-as-bid \cite{eso2021ancillary,eso2021balancing}.} Although advanced mechanisms, which limit market manipulations and allow fairer welfare distribution between participants, are proposed in the literature, in practice, the conventional mechanisms are being adopted to demand-side flexibility services \cite{ukpowernetworks2021}. 
\textcolor{black}{Furthermore, DSO services are a new concept and the suitability of existing market mechanisms cannot be assumed, particularly as the limited number of available participants within the distribution network can reduce the level of competition within the market.}
In either case, before a mechanism can be applied to a flexibility service, it is essential to have a framework to enable comprehensive analysis and comparison of the efficacy of market mechanisms.

\textcolor{black}{In the existing literature, it is assumed that flexibility service providers deliver the service at an actual cost (i.e., provide truthful offers), while the possibility to amend their offers is omitted. We find this aspect to be very critical for DSO service planning, because strategic bidding will directly affect prices for the flexibility service and DSO costs.} 
The previous research on the whole system markets and services showed that there is a significant value in bidding opportunities when it comes to a particular market mechanism, which have been illustrated for wholesale electricity markets by Bakirtzis \textit{et al.} in \cite{bakirtzis2006agent} and by Chirkin \textit{et al.} in \cite{chirkin2016gaming}. In \cite{poplavskaya2020effect}, Poplavskaya \textit{et al.} have studied the effect of market design on strategic behaviour of service providers for the European balancing markets. Analysis of the literature suggests there is a lack of research on strategic behaviour of stakeholders, bidding opportunities, and market manipulations for demand-side flexibility.

There are also very few studies that test the emerging demand-side flexibility concept on a real case study. While it is important to illustrate that the proposed methods and algorithms can be implemented for a \textcolor{black}{benchmark system}, it is essential to test the new approaches in a real case study to estimate applicability and the expected benefits of the new concept.

\subsection{Scope of the paper and contributions}
This paper provides a two-step optimization-based modelling and game-theoretic framework to assess the capability of demand-side flexibility to provide services to a local distribution network. The framework can be used to investigate the consequence of different market mechanisms -- including their susceptibility to market manipulations -- and is demonstrated using a real-world case study. 
In the first step, optimization-based models using a generalised problem formulation are applied to determine available flexible capacity from different technologies as a function of their marginal cost. These functions are based on realistic modelling of flexible assets, their technical constraints, customer preferences and cost of energy. The analysis yields realistic demand-side flexibility offer curves for numerous flexible assets, i.e., HPs, EVs, EES, and industrial and commercial demand-side response (I\&C DSR). In the second step, the obtained offer curves are used as inputs to an MA iterative game, which determines expected flexibility service market equilibrium prices and capacity traded for various market clearing mechanisms. In this MA modelling method, flexibility providers follow strategic bidding algorithms in an iterative manner until an equilibrium state is found. Three bidding strategies are studied: overpricing, capacity understatement, and underbidding. Each strategy is applied to four market clearing mechanisms: pay-as-bid (PAB), pay-as-cleared (PAC), Dutch reverse auction (DRA), and Vickrey-Clarke-Groves (VCG). 
The resulting two-step framework comprehensively accounts for the individual objectives of flexibility providers, technical constraints of flexible assets, customer preferences, market clearing, and strategic bidding.

The proposed methodology is demonstrated using a real case study, in which local DNO is interested in investigating technical potential and economic effect of a prospective demand-side flexibility service as part of the Customer-Led Distribution System project \cite{clds2021}.
\textcolor{black}{For the cases considered,} the numerical study showed that in three out of the four \textcolor{black}{customer engagement} scenarios, there is enough capacity to meet DSO flexibility requirements. However, depending on the various factors, i.e., competition level and market mechanisms, the expected price for flexibility service can vary by as much as 87.5\% compared to the minimum found \textcolor{black}{due to the bidding. It has been observed that the strategy that will provide service providers with the highest payoff (i.e., dominant strategy) varies with market clearing mechanisms. For instance, for the PAC, the dominant strategy for service providers is a capacity understatement; for the PAB, it is overpricing; for the VCG, it is providing truthful offers, which have been known previously and observed in our results. However, for the considered demand-side flexibility service procured through the DRA, overpricing strategy is found to be more effective for service providers than underbidding, while the former was previously reported as a dominant strategy for the DRA in \cite{gillen2016bid}; as such, this constitutes a novel finding of this work. Finally, given the vulnerability of the PAB and DRA to manipulations, it has been concluded} that the uniform pricing mechanism, such as PAC, provides limited opportunities for manipulations at contracted flexibility market even when competition is low, while the discriminatory pricing mechanisms, such as PAB and DRA, are prone to manipulations, especially, in the lack of competition, resulting in a significantly higher price for the flexibility service. \textcolor{black}{Crucially, our approach enables evaluation of many other case-specific metrics for DSO planning, which we also provide in the results. These include the amount of flexible capacities from flexible assets and their marginal costs (i.e., supply curves), expected prices for each market clearing mechanism, including their ranges and average values, costs and benefits for the DSO, and costs and profits for flexibility service providers.}

To summarize, the main contributions of the present paper are:
\begin{enumerate}
    \item A generalized optimization problem formulation is proposed to determine flexibility supply and its marginal cost from demand-side flexibility providers, accounting for the parameters of the flexible assets, customer preferences, and price for energy.
    \item Realistic flexibility supply functions, i.e., capacity versus marginal cost, are determined for various flexible assets using the developed generic optimization problem formulation that accounts for realistic models of flexible assets, their technical constraints, customer preferences, and ToU tariff.
    \item A game-theoretic approach, which accounts for strategic behaviour of market participants, is developed to determine the expected costs and benefits of contracted flexibility for the DSO and demand-side flexibility providers.
    \item The efficacy of existing market mechanisms for a prospective DSO flexibility service is evaluated, considering a real GB case study of residential sector in the North East of England.
    \item \textcolor{black}{The dominant strategy for the considered demand-side flexibility service procured through the DRA is found to be overpricing (in contrast to the existing literature \cite{gillen2016bid}). This reveals vulnerability of the DRA mechanism to market manipulations that leads to the significant increase of the average equilibrium price and similar results to PAB.}
\end{enumerate}

\section{Step I - Modelling distributed flexibility}\label{sec:flex_models}

This section describes the optimization-based modelling approach used to determine the offer curve of demand-side flexibility providers in terms of available flexible capacity and its marginal cost. These curves are a key input to the game-theoretic framework in Step II. The structure of the proposed optimization problems accounts for the objectives of demand-side flexibility providers, operation costs, customer preferences, and technical constraints of flexible assets. The general formulation is:
\begin{equation}\label{eq:gen_obj}
	\begin{gathered}
		\max_{P^\text{F},P_t^\text{Sch},\bm{x}} [\pi P^\text{F} \Delta \text{T}^\text{FW} - Cost(P_t^\text{Sch},\bm{x}) - Pen(P_t^\text{Sch},\bm{x})]
	\end{gathered}
\end{equation}
subject to
\begin{equation}\label{eq:gen_con1}
	h_m (P_t^\text{Sch},\bm{x}) \leq 0, \:\: m \in M,
\end{equation}
\begin{equation}\label{eq:gen_con2}
	P^\text{F} \leq \text{P}_t^\text{Org} - P_t^\text{Sch} \:\: \forall t \in T^{FW},
\end{equation}
where the objective function \eqref{eq:gen_obj} is to be maximized with respect to flexible capacity $P^\text{F} \in \mathbb{R}_{\geq 0}$, scheduled power consumption/generation of flexible assets $P_t^\text{Sch} \in \mathbb{R}$, and other variables $\bm{x} \in \mathbb{R}^n$ that describe decision variables related to flexible devices. Offer curves can be obtained by solving the optimization problem for different values of $\pi$ and determining $P^\text{F}(\pi)$ - the maximum flexibility available at that price.

In the objective function \eqref{eq:gen_obj}, the reward for the contracted flexibility service is calculated as a payment for availability of the minimum flexible capacity $P^\text{F}$ available during the flexibility time window $T^\text{FW}$, where $\pi$ is availability fee and $\Delta \text{T}^\text{FW}$ is the duration of flexibility time window. The operating costs of flexible assets are formulated in $Cost(P_t^\text{Sch},\bm{x})$, while penalties are in $Pen(P_t^\text{Sch},\bm{x})$. The latter can correspond to the service nondelivery or the customer discomfort associated with the service delivery. The set of constraints \eqref{eq:gen_con1} includes models and technical constraints of flexible assets, where $M$ is a set of constraints indexed by $m$. Constraint \eqref{eq:gen_con2} defines flexible capacity that can be delivered by demand-side flexibility. The latter is found with respect to $\text{P}_t^\text{Org}$, which can be given as the original demand consumption of flexible assets or scheduled consumption corresponding to the solution of problem \eqref{eq:gen_obj}-\eqref{eq:gen_con1}, when $\pi = 0$.

\subsection{Flexibility from heat pumps}

The optimization problem to determine demand-side flexibility from domestic heating is formulated in \eqref{eq:hp_obj}-\eqref{eq:hp_con5}: 
\begin{equation}\label{eq:hp_obj}
    \begin{gathered}
     \max_{P^\text{F},P_{d,t}^\text{HP},\tau_{d,t},\tau_{d,t}^\text{P},\tau_{d,t}^\text{N}} [\pi P^\text{F} \Delta \text{T}^\text{FW} - \text{N}^\text{HP} \sum_{d \in D} \mu_d \cdot \\ \cdot \sum_{t \in T} (\text{c}_t^\text{En} P_{d,t}^\text{HP} \Delta t + \frac{\text{c}^\text{Pen}}{2} ({\tau_{d,t}^\text{P}}^2 + {\tau_{d,t}^\text{N}}^2) \Delta t)]
    \end{gathered}
\end{equation}
subject to
\begin{equation}\label{eq:hp_con1}
   0 \leq P_{d,t}^\text{HP} \leq \bar{\text{P}}_d^\text{HP} \:\: \forall d \in D, t \in T,
\end{equation}
\begin{equation}\label{eq:hp_con2}
    \begin{gathered}
       \tau_{d,t+1} = \tau_{d,t} - \frac{\text{K}_d^\text{Th}}{\text{C}_d^\text{Th}}(\tau_{d,t} - \tau_t^\text{Amb}) \Delta t + \frac{\text{k}^\text{HP}}{\text{C}_d^\text{Th}} P_{d,t}^\text{HP} \Delta t \:\: \forall d \in D, t \in T,
   \end{gathered}
\end{equation}
\begin{equation}\label{eq:hp_con3}
   \tau_{d,1} = \tau_{d,\text{T}+1} \:\: \forall d \in D,
\end{equation}
\begin{equation}\label{eq:hp_con4}
   P_{d,t}^\text{HP} \leq \frac{\text{p}^\text{HP}}{\Delta \text{T}} \sum_{t' \in T} P_{d,t'}^\text{HP} \:\: \forall d \in D, t \in T,
\end{equation}
\begin{equation}\label{eq:hp_con6}
   \tau_{d,t}^\text{P} >= \tau_{d,t} - \tau^\text{Max} \:\: \forall d \in D, t \in T,
\end{equation}
\begin{equation}\label{eq:hp_con7}
   \tau_{d,t}^\text{N} >= \tau^\text{Min} - \tau_{d,t} \:\: \forall d \in D, t \in T,
\end{equation}
\begin{equation}\label{eq:hp_con5}
 \begin{gathered}
   P^\text{F} \leq \text{N}^\text{HP} \sum_{d \in D} \mu_d (\frac{\text{K}_d^\text{Th}}{\text{k}^\text{HP}}(\frac{\tau^\text{Max} + \tau^\text{Min}}{2} - \tau_t^\text{Amb}) - P_{d,t}^\text{HP}) \:\: \forall t \in {T}^\text{FW},
 \end{gathered}
\end{equation}
where the optimization problem variables include flexible capacity $P^\text{F} \in \mathbb{R}_{\geq 0}$, heating demand $P_{d,t}^\text{HP} \in \mathbb{R}_{\geq 0}$, indoor temperature $\tau_{d,t} \in \mathbb{R}_{\geq 0}$, and auxiliary variables $\tau_{d,t}^\text{P} \in \mathbb{R}_{\geq 0}$ and $\tau_{d,t}^\text{N} \in \mathbb{R}_{\geq 0}$ that represent deviation from the comfortable indoor temperature range [$\tau^\text{Min}$,$\tau^\text{Max}$]. $D$ is a set of dwelling types indexed by $d$, $T$ is a set of time intervals indexed by $t$ with a time step $\Delta t$.

In the objective function (\ref{eq:hp_obj}), the first term determines the reward for available flexible capacity as per the generalized formulation. The second term formulates operation costs of flexible assets, where $\text{N}^\text{HP}$ is the number of households equipped with HPs, $\mu_d$ is the share of each dwelling type, and $\text{c}_t^\text{En}$ is a ToU tariff. The third term represents the penalty for deviation of indoor temperature $\tau_{d,t}$ outside the comfortable indoor temperature range, where $\text{c}^\text{Pen}$ represents specific value of customer discomfort. In the constraints of the optimization problem above, \eqref{eq:hp_con1} limits the heating demand $P_{d,t}^\text{HP}$ to the device power rating $\bar{\text{P}}_d^\text{HP}$. The second constraint \eqref{eq:hp_con2} formulates a linear thermal model of dwellings, where $\text{K}_d^\text{Th}$ is an aggregate thermal conductance of walls, windows and entrance doors, $\text{C}_d^\text{Th}$ is the thermal capacitance of dwellings, $\tau_t^\text{Amb}$ is the ambient temperature profile, and $\text{k}^\text{HP}$ is a heating device conversion factor (e.g., for HPs the value of $\text{k}^\text{HP}$ may vary from three to four, while for resistive heating it is equal to one). Constraint \eqref{eq:hp_con3} establishes that the final temperature value must equal the initial temperature value. Constraint \eqref{eq:hp_con4} limits the peak demand from heating by the factor of $\text{p}^\text{HP}$ from the average daily heating demand, which is needed to avoid uncontained peak demand just before the flexibility time window. Constraints \eqref{eq:hp_con6} and \eqref{eq:hp_con7} determine the deviation of indoor temperature from the comfortable range. The final constraint \eqref{eq:hp_con5} defines the available demand-side flexibility, which is found as the difference between the expected and dispatched electricity demand for heating.

\subsection{Flexibility from EV charging}

The optimization problem to determine demand-side flexibility from EV charging is formulated in \eqref{eq:ev_obj}-\eqref{eq:ev_con6}:
\begin{equation}\label{eq:ev_obj}
     \max_{P^\text{F},E_t^\text{EV},P_t^\text{EVT},P_t^\text{EVB},E_t^\text{EVL}} [\pi P^\text{F} \Delta \text{T}^\text{FW} - \sum_{t \in T} (\text{c}_t^\text{En} P_{t}^\text{EVT} \Delta t 
        + \frac{\text{c}^\text{Pen}}{2} {E_t^\text{EVL}}^2 \Delta t)]
\end{equation}
subject to
\begin{equation}\label{eq:ev_con1}
   0 \leq P_t^\text{EVT} \leq \frac{\text{N}_t^\text{Plg}}{\text{N}^\text{EV}} \bar{\text{P}}^\text{EV} \:\: \forall t \in T,
\end{equation}
\begin{equation}\label{eq:ev_con2}
   P_t^\text{EVT} \geq P_t^\text{EVB} + \frac{\text{R}^\text{In}}{{\text{V}^\text{OC}}^2}{P_t^\text{EVB}}^2 \:\: \forall t \in T,
\end{equation}
\begin{equation}\label{eq:ev_con3}
   E_{t+1}^\text{EV} = E_t^\text{EV} + P_t^\text{EVB} \Delta t \:\: \forall t \in T,
\end{equation}
\begin{equation}\label{eq:ev_con4}
   E_{\text{T}+1}^\text{EV} = E_1^\text{EV},
\end{equation}
\begin{equation}\label{eq:ev_con5}
   E_t^\text{EVL} \geq \text{e}^\text{EV}(1 - \frac{\text{N}_t^\text{Plg}}{\text{N}^\text{EV}}) - E_t^\text{EV} \: \forall t \in [2:00;14:00],
\end{equation}
\begin{equation}\label{eq:ev_con6}
   P^\text{F} \leq \text{P}_t^\text{EVDem} - P_t^\text{EVT} \:\: \forall t \in T^\text{FW},
\end{equation}
where the optimization problem variables include flexible capacity $P^\text{F} \in \mathbb{R}_{\geq 0}$, state of energy of the aggregation of EV vehicles $E_t^\text{EV} \in \mathbb{R}_{\geq 0}$, charging power at terminals $P_t^\text{EVT} \in \mathbb{R}_{\geq 0}$ and at the battery $P_t^\text{EVB} \in \mathbb{R}_{\geq 0}$, the energy demand that has not been met in time $E_t^\text{EVL} \in \mathbb{R}_{\geq 0}$.

The objective function \eqref{eq:ev_obj}, comprises the reward for available flexibility minus EVs charging costs and a penalty term, which accounts for the non-delivery of EV charging demand. Constraint \eqref{eq:ev_con1} defines the available charging capacity based on the relative number of plugged in EVs, where $\bar{\text{P}}^\text{EV}$ is an aggregate capacity of EV chargers, $\text{N}_t^\text{Plg}$ is a number of plugged in EVs, while $\text{N}^\text{EV}$ is the total number of EVs. 
The second constraint \eqref{eq:ev_con2} is a convex relaxation of the battery Rint model \cite{sayfutdinov2021optimal} that models charging efficiency, where $\text{R}^\text{In}$ is the battery internal resistance and $\text{V}^\text{OC}$ is the battery open-circuit voltage. It is worth noting that the exactness of \eqref{eq:ev_con2} is achieved when the constraint is binding. The latter is ensured when the objective function is aimed to cost minimization.
Constraint \eqref{eq:ev_con3} constitutes continuity of the charged energy, while the next constraint \eqref{eq:ev_con4} establishes zero daily net charge. Constraint \eqref{eq:ev_con5} defines the amount of charge that has not been met in time, where $\text{e}^\text{EV}$ is the total daily energy demand from the aggregation of EVs. In \eqref{eq:ev_con5}, we assume that EVs need to be charged by the time they leave, which is normally between 2am and 2pm, with the majority of the EVs having departed by 10am \cite{madzharov2014integrating}. The last constraint \eqref{eq:ev_con6} defines flexible capacity that can be delivered by demand-side flexibility, which is found as a difference between the uncontrolled EV charging demand $\text{P}_t^\text{EVDem}$ and the charging demand $P_t^\text{EVT}$.

\subsection{Flexibility from EES}

The optimization problem to determine demand-side flexibility from EES is formulated in \eqref{eq:es_obj}-\eqref{eq:es_con9}:
\begin{equation}\label{eq:es_obj}
    \begin{gathered}
     \max_{P^\text{F},E_t^\text{ES},P_t^\text{Ch},P_t^\text{Dis},DoD,\alpha_i} [\pi P^\text{F} \Delta \text{T}^\text{FW} - \\ - \sum_{t \in T} \text{c}_t^\text{En} (P_{t}^\text{Ch} + P_{t}^\text{Dis}) \Delta t - \frac{\bar{\text{E}} \: \text{C}^\text{E}}{\sum_{i \in I} \alpha_i \text{N}_i^\text{Cyc}}]
    \end{gathered}
\end{equation}
subject to
\begin{equation}\label{eq:es_con1}
   0 \leq P_t^\text{Ch} \leq \bar{\text{P}}^\text{ES} \:\: \forall t \in T,
\end{equation}
\begin{equation}\label{eq:es_con2}
   -\bar{\text{P}}^\text{ES} \leq P_t^\text{Dis} \leq 0 \:\: \forall t \in T,
\end{equation}
\begin{equation}\label{eq:es_con3}
   E_{t+1}^\text{ES} = E_t^\text{ES} + (\eta^\text{Ch} P_t^\text{Ch} + \frac{P_t^\text{Dis}}{\eta^\text{Dis}}) \Delta t \:\: \forall t \in T,
\end{equation}
\begin{equation}\label{eq:es_con4}
   E_{\text{T}+1}^\text{ES} = E_1^\text{ES},
\end{equation}
\begin{equation}\label{eq:es_con5}
   0 \leq E_t^\text{ES} \leq \bar{\text{E}} \:\: \forall t \in T,
\end{equation}
\begin{equation}\label{eq:es_con6}
   DoD = \frac{1}{2 \: \bar{\text{E}}} \sum_{t \in T}(\eta^\text{Ch} P_t^\text{Ch} - \frac{P_t^\text{Dis}}{\eta^\text{Dis}}) \Delta t,
\end{equation}
\begin{equation}\label{eq:es_con7}
   \sum_{i \in I} \underline{\text{DoD}}_i \: \alpha_i \leq DoD \leq \sum_{i \in I} \overline{\text{DoD}}_i \: \alpha_i,
\end{equation}
\begin{equation}\label{eq:es_con8}
   \sum_{i \in I} \alpha_i = 1,
\end{equation}
\begin{equation}\label{eq:es_con9}
   P^\text{F} \leq -(P_t^\text{Ch} + P_t^\text{Dis}) \:\: \forall t \in {T}^\text{FW},
\end{equation}
where the optimization problem variables include flexible capacity $P^\text{F} \in \mathbb{R}_{\geq 0}$, cumulative energy storage state of energy $E_t^\text{ES} \in \mathbb{R}_{\geq 0}$, charging $P_t^\text{Ch} \in \mathbb{R}_{\geq 0}$ and discharging $P_t^\text{Dis} \in \mathbb{R}_{\leq 0}$ power, depth of discharge of energy storage $DoD \in \mathbb{R}_{[0,1]}$, auxiliary binary variables $\alpha_i \in \mathbb{Z}_{[0,1]}$.

The objective function \eqref{eq:es_obj} includes the reward and costs terms, where the latter accounts for the energy storage operation costs, i.e., charging/ discharging, cycling, and cost of energy. There is no penalty term associated with EES. This formulation employs a Depth of Discharge (DoD)-based cycle lifetime model using Mixed-Integer Linear Programming (MILP), where $\text{N}_i^\text{Cyc}$ provides cycle lifetime that corresponds to the $i$-th range of DoD, indicated by the auxiliary variable $\alpha_i$. Constraints \eqref{eq:es_con1} and \eqref{eq:es_con2} limit the power output to the installed power capacity $\bar{\text{P}}^\text{ES}$. Constraint \eqref{eq:es_con3} constitutes energy storage continuity, where $\eta^\text{Ch}$ and $\eta^\text{Dis}$ are charge and discharge efficiencies, respectively. Net zero charge is ensured by \eqref{eq:es_con4}, while the storage capacity limits are respected in \eqref{eq:es_con5}. Energy storage DoD is determined in \eqref{eq:es_con6}, and a particular range of DoD is found in \eqref{eq:es_con7} and \eqref{eq:es_con8}, where $\underline{\text{DoD}}_i$ and $\overline{\text{DoD}}_i$ are the low and high bounds of the linear segment $i$. Finally, the last constraint \eqref{eq:es_con9} defines flexible capacity that can be delivered, which is found as a negative sum of charging and discharging power.

\subsection{Flexibility from I\&C DSR}

The optimization problem to determine demand-side flexibility from I\&C DSR is formulated in \eqref{eq:ic_obj}-\eqref{eq:ic_con3}:
\begin{equation}\label{eq:ic_obj}
	\begin{gathered}
		\max_{P^\text{F},P_t^\text{I\&C}} [\pi P^\text{F} \Delta \text{T}^\text{FW} - C^\text{I\&C}(P^\text{F}) - \sum_{t \in T} \text{c}_t^\text{En} P_{t}^\text{I\&C} \Delta t]
	\end{gathered}
\end{equation}
subject to
\begin{equation}\label{eq:ic_con1}
 	P^\text{F} \Delta \text{T}^\text{FW} = \text{e}^\text{I\&C} \sum_{t \in T^\text{Rec}} P_t^\text{I\&C} \Delta t,
\end{equation}
\begin{equation}\label{eq:ic_con2}
	P_t^\text{I\&C} \leq \text{p}^\text{I\&C} P^\text{F} \:\: \forall \: t \in T^\text{Rec},
\end{equation}
\begin{equation}\label{eq:ic_con3}
	P^\text{F} \leq \bar{\text{P}}^\text{I\&C} - P_t^\text{I\&C} \:\: \forall \: t \in T^\text{FW},
\end{equation}
where the problem variables include flexible capacity $P^\text{F} \in \mathbb{R}_{\geq 0}$, and I\&C flexible power consumption $P_t^\text{I\&C} \in \mathbb{R}_{\geq 0}$.

The objective function \eqref{eq:ic_obj} includes the reward and operation costs, with the latter comprising the I\&C DSR cost function and cost of energy. The first constraint \eqref{eq:ic_con1} establishes the requirement for DSR load recovery, where $\text{e}^\text{I\&C}$ is an energy recovery factor that determines percentage of DSR energy that should be recovered, $T^\text{Rec}$ is a load recovery time period that usually occurs right after a DSR event, i.e., flexibility time window. The second constraint \eqref{eq:ic_con2} limits the I\&C flexible power consumption by the power recovery factor $\text{p}^\text{I\&C}$ for the load recovery time period. Finally, the last constraint \eqref{eq:ic_con3} determines the amount of flexible capacity.



\section{Step II - Multi-agent modelling framework}\label{sec:ma_framework}

The MA framework models the interactions of flexibility providers, i.e., agents, with DSO service market, in which agents are assumed to follow a particular bidding strategy to maximize their profit. 

\subsection{Multi-agent framework}

\begin{figure}
\centering
        \includegraphics[width=0.5\textwidth]{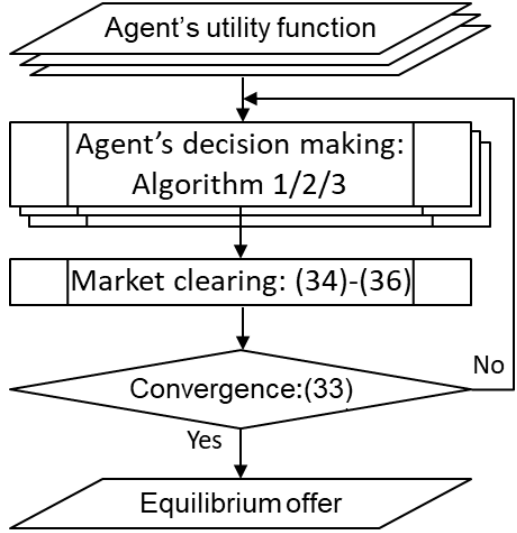}
        \textcolor{black}{\caption{Multi-agent framework}}
        \label{fig:ma_frame}
\end{figure}

MA modelling is implemented using iterative game approach illustrated in Fig.~1. In this framework, agents interact with the DSO market in a repeated manner, where a decision of each agent is based on its utility function and the outcomes of the previous round of market clearing. Iterative games are repeated until the agents stop changing their offers, i.e., capacity and price pairs. The resulting offers, obtained in the last iteration, are considered equilibrium. This equilibrium forms the convergence criterion for the termination of MA modelling, and is formulated as:

\begin{equation}\label{eq:ma_conv}
   \lVert \bm{\Delta \pi} \rVert_2 + \lVert \bm{\Delta P} \rVert_2 \leq \epsilon,
\end{equation}
where $\bm{\Delta \pi} = \{\Delta \pi_a \in \mathbb{R}\}$ is the change in offer price, $\bm{\Delta P} = \{\Delta P_a \in \mathbb{R}\}$ is the change in offered capacity\textcolor{black}{, and $\epsilon$ is a tolerance value}. Index $a \in A$ denotes a particular agent.


\subsection{Market clearing}

The market clearing process, illustrated in Fig.~1 is formulated as follows:
\begin{equation}\label{eq:market_obj}
    \min_{\bm{P}^\textbf{A},x} \bm{P}^\textbf{A} {\bm{\pi}^\textbf{O}}^\text{T} + x \bar{\pi}
\end{equation}
subject to
\begin{equation}\label{eq:market_con1}
   0 \leq \bm{P}^\textbf{A} \leq \textbf{P}^\textbf{O},
\end{equation}
\begin{equation}\label{eq:market_con2}
   \text{P}^\text{D} = \lVert \bm{P}^\textbf{A} \rVert_1 + x \: : \: \lambda,
\end{equation}
where the optimization problem matches flexibility demand $\text{P}^\text{D} \in \mathbb{R}_{\geq 0}$ with offers $\textbf{P}^\textbf{O} = \{\textbf{P}^\textbf{O}_a \in \mathbb{R}_{\geq 0}^{|G|}\}$ at the least cost, where $G=\{1,2,..,\overline{\text{G}}\}$ is a set of price levels and $|G|$ is its cardinality. The optimization problem \eqref{eq:market_obj}-\eqref{eq:market_con2} is solved with respect to a vector of accepted offers from all agents $\bm{P}^\textbf{A} = \{\bm{P}^\textbf{A}_a \in \mathbb{R}_{\geq 0}^{|G|}\}$ and auxiliary variable $x \in \mathbb{R}_{[0,\text{P}^\text{D}]}$ that represents any unmet flexibility demand.

In the objective function (\ref{eq:market_obj}), the first term constitutes the costs for accepted flexible capacity $\bm{P}^\textbf{A}$ with offer price $\bm{\pi}^\textbf{O} = \{ \bm{\pi}_a^\textbf{O} \in \mathbb{R}_{\geq 0}^{|G|}\}$, while the second term limits the acceptance of offers, excluding those whose price exceeds ceiling price $\bar{\pi}$. The first constraint \eqref{eq:market_con1} limits accepted capacity $\bm{P}^\textbf{A}$ to the offered capacity $\textbf{P}^\textbf{O}$, while the second constraint \eqref{eq:market_con2} establishes that the demand capacity is equal to the accepted capacity plus the unmet demand.

The inputs of the optimization problem \eqref{eq:market_obj}-\eqref{eq:market_con2} are flexibility demand $\text{P}^\text{D}$ and ceiling price $\bar{\pi}$ issued by DSO, and pairs of flexible capacity $\textbf{P}^\textbf{O}_a$ and price $\bm{\pi}^\textbf{O}_a$ offered by each agent $a$. Depending on the market clearing mechanism applied, the outputs of the optimization problem may vary. For the discriminatory price markets, i.e., PAB and DRA, the output includes only accepted capacity $\bm{P}^\textbf{A}_a$, which is rewarded by the offer price $\bm{\pi}^\textbf{O}_a$, known privately to each agent. In case of the uniform price market, i.e., PAC, the price is uniform for all accepted offers, and in addition to the accepted capacity $\bm{P}^\textbf{A}_a$ the agents are informed of the Market Clearing Price (MCP), which is found as a dual variable $\lambda$ of constraint \eqref{eq:market_con2}.

\subsection{Agent's decision-making}

This section presents three strategic bidding algorithms -- overpricing, understatement, and underbidding -- which can be followed by the agents to improve their profit as a result of market conditions and the market clearing mechanism. The strategic bidding algorithms are provided in Algorithms~\ref{alg:OP}-\ref{alg:UB}.

The overpricing (OP) strategy (Algorithm~\ref{alg:OP}) forces the agents offer their maximum capacity values $\textbf{P}^\textbf{O}_a = \textbf{P}^\textbf{True}_a$ 
, while increasing the offer prices $\bm{\pi}^\textbf{O}_a$ above the true marginal cost $\bm{\pi}_a^\textbf{True}$ to gain profit.
It should be noted that the vectors $\textbf{P}^\textbf{True}_a$ and $\bm{\pi}_a^\textbf{True}$ form the flexibility supply curve, which is found by the optimization-based modelling approach described in the previous section.
The profit of an agent on the $i$-th game iteration is computed in step 2, where $\bm{E}$ is a vector of ones. The decision-making is formulated in the if-else blocks, where an agent decides on its offer price vector $\bm{\pi}^{\textbf{O}(i+1)}_a$ for the next round $(i+1)$ based on the two logical conditions. The first condition (step 3) $\Pi_a^{(i)} \geq \Pi_a^{(i-1)}$ determines if the current profit is greater or equal to the one obtained in the previous iteration, while the second condition (steps 4 and 12) $\Delta \pi^{(i-1)}_a \geq 0$ determines the direction of the last price update that caused change in the profit. For instance, if an agent increased its offer prices in the last iteration and this led to an increase in profit, this agent will increase its offer prices again by the same amount in the next round (steps 5 and 6). However, if an agent increased its offer prices in the previous iteration and this led to a reduction in profit, it will decrease its offer prices in the next round (steps 13 and 14). The step size is halved and the offer prices are increased (steps 8 and 9) if the profit increase was a result of offer price reduction.

\begin{algorithm}
    \caption{Overpricing strategy}
    \label{alg:OP}
  \begin{algorithmic}[1]
    \REQUIRE Decide on the offer price
    \INPUT $\bm{P}^{\textbf{A}(i)}_a$; $\lambda^{(i)}$ 
    \OUTPUT $\textbf{P}^\textbf{O}_a$; $\bm{\pi}^{\textbf{O}(i+1)}_a$
    \STATE \textbf{Init.:} \: $\textbf{P}^\textbf{O}_a = \textbf{P}^\textbf{True}_a$; ${\pi}^{\text{O}(1)}_{a,g} = \lambda^{(1)} \: \forall g \in G | {\pi}_{a,g}^\text{True} \leq \lambda^{(1)}$; $\Pi_a^{(0)} = 0$
    
    \STATE $\Pi_a^{(i)} =  \begin{cases}
    	\bm{P}^{\textbf{A}(i)}_a (\bm{E} \cdot \lambda^{(i)} - \bm{\pi}_a^\textbf{True})^T, \text{for uniform pricing;} \\
    	\bm{P}^{\textbf{A}(i)}_a (\bm{\pi}^{\textbf{O}(i)}_a - \bm{\pi}_a^\textbf{True})^T, \text{for discriminatory pricing.}
    \end{cases}$
            
        \IF{$\Pi_a^{(i)} \geq \Pi_a^{(i-1)}$}
            \IF{$\Delta \pi^{(i-1)}_a \geq 0$}
                \STATE $\Delta \pi^{(i)}_a = \Delta \pi^{(i-1)}_a$
                \STATE $\pi^{\text{O}(i+1)}_{a,g} = \pi^{\text{O}(i)}_{a,1} + \Delta \pi^{(i)}_a \forall g \in G | {\pi}_{a,g}^\text{True} \leq \pi^{\text{O}(i)}_{a,1} + \Delta \pi^{(i)}_a$
            \ELSE
                \STATE $\Delta \pi^{(i)}_a = \Delta \pi^{(i-1)}_a / 2$
                \STATE $\pi^{\text{O}(i+1)}_{a,g} = \pi^{\text{O}(i)}_{a,1} + \Delta \pi^{(i)}_a \forall g \in G | {\pi}_{a,g}^\text{True} \leq \pi^{\text{O}(i)}_{a,1} + \Delta \pi^{(i)}_a$
                \ENDIF
        \ELSE
           \IF{$\Delta \pi^{(i-1)}_a \geq 0$}
                \STATE $\Delta \pi^{(i)}_a = - \Delta \pi^{(i-1)}_a$
                \STATE $\pi^{\text{O}(i+1)}_{a,g} = \pi^{\text{O}(i)}_{a,1} + \Delta \pi^{(i)}_a \forall g \in G | {\pi}_{a,g}^\text{True} \leq \pi^{\text{O}(i)}_{a,1} + \Delta \pi^{(i)}_a$
            \ELSE
                \STATE $\Delta \pi^{(i)}_a = \Delta \pi^{(i-1)}_a$ 
                \STATE $\pi^{\text{O}(i+1)}_{a,g} = \pi^{\text{O}(i)}_{a,1} + \Delta \pi^{(i)}_a \forall g \in G | {\pi}_{a,g}^\text{True} \leq \pi^{\text{O}(i)}_{a,1} + \Delta \pi^{(i)}_a$
            \ENDIF
        \ENDIF
  \end{algorithmic}
\end{algorithm}

An understatement (US) strategy -- which is presented in Algorithm~\ref{alg:US} -- implies that the agents understate their offer capacity $\textbf{P}^\textbf{O}_a$, while keeping the true offer price $\bm{\pi}^\textbf{O}_a = \bm{\pi}_a^\textbf{True}$. The rationale behind the strategy is that agents can benefit from capacity understatement by forcing more expensive offers to be accepted. According to the algorithm, the decision-making is performed at step 3, in which an agent decides to decrease its capacity offers if the current profit is higher or equal to the profit obtained in the previous iteration. Otherwise, an agent decreases its step $\Delta P^{(i)}_a$ twice and increases its capacity offer accordingly.

\begin{algorithm}
    \caption{Understatement strategy}
    \label{alg:US}
  \begin{algorithmic}[1]
    \REQUIRE Decide on the offer capacity
    \INPUT $\bm{P}^{\textbf{A}(i)}_a$; $\lambda^{(i)}$ 
    \OUTPUT $\textbf{P}^{\textbf{O}(i+1)}_a$; $\bm{\pi}^\textbf{O}_a$
    \STATE \textbf{Initialization} \: $\textbf{P}^{\textbf{O}(1)}_a = \textbf{P}^\textbf{True}_a$; $\bm{\pi}^\textbf{O}_a = \bm{\pi}_a^\textbf{True}$, $\Pi_a^{(0)} : = 0$

    \STATE $\Pi_a^{(i)} =  \begin{cases}
            \bm{P}^{\textbf{A}(i)}_a (\bm{E} \cdot \lambda^{(i)} - \bm{\pi}_a^\textbf{True})^T, \text{for uniform pricing;} \\
            \bm{P}^{\textbf{A}(i)}_a (\bm{\pi}^{\textbf{O}}_a - \bm{\pi}_a^\textbf{True})^T, \text{for discriminatory pricing.}
                \end{cases}$

    \IF{$\Pi_a^{(i)} \geq \Pi_a^{(i-1)}$}
        \STATE $\Delta P^{(i)}_a = \Delta P^{(i-1)}_a$
        \STATE $\text{P}^{\text{O}(i+1)}_{a,g} = \text{P}^{\text{O}(i)}_{a,g} - \Delta P^{(i)}_a \: \forall g \in G | \pi^{\text{O}}_{a,g} = \lambda^{(i)}$
    \ELSE
        \STATE $\Delta P^{(i)}_a = \Delta P^{(i-1)}_a / 2$
        \STATE $\text{P}^{\text{O}(i+1)}_{a,g} = \text{P}^{\text{O}(i)}_{a,g} + \Delta P^{(i)}_a \: \forall g \in G | \pi^{\text{O}}_{a,g} = \lambda^{(i)}$
    \ENDIF
  \end{algorithmic}
\end{algorithm}

\begin{algorithm}
    \caption{Underbidding strategy}
    \label{alg:UB}
  \begin{algorithmic}[1]
    \REQUIRE Decide on the offer price
    \INPUT $\bm{P}^{\textbf{A}(i)}_a$; $\textbf{P}^{\textbf{DM}(i)}$; $\bm{\pi}^{\textbf{DM}(i)}$; $\text{P}^{\text{D}}$
    \OUTPUT $\textbf{P}^\textbf{O}_a$; $\bm{\pi}^{\textbf{O}(i+1)}_a$
    \STATE \textbf{Init.:} \: $\textbf{P}^\textbf{O}_a = \textbf{P}^\textbf{True}_a$; ${\pi}^{\text{O}(1)}_{a,g} = \pi^\text{Start}_a \: \forall g \in G | {\pi}_{a,g}^\text{True} \leq \pi^\text{Start}$

    \STATE $\Pi_a^{(i)} =  \begin{cases}
    	\bm{P}^{\textbf{A}(i)}_a (\bm{E} \cdot \lambda^{(i)} - \bm{\pi}_a^\textbf{True})^T, \text{for uniform pricing;} \\
    	\bm{P}^{\textbf{A}(i)}_a (\bm{\pi}^{\textbf{O}(i)}_a - \bm{\pi}_a^\textbf{True})^T, \text{for discriminatory pricing.}
    \end{cases}$
    
    \STATE $\Pi_a^{\text{UB}(i)} = \sum_{g \in G} [min\{\text{P}^{\text{D}} - \text{P}^{\text{DM}(i)}_g, \text{P}_{a,g}^{\text{O}}\} \cdot ({\pi}^{\text{DM}(i)}_{g-1} - \pi_{a,g}^{\text{True}})]$
    
    \IF{$\Pi_a^{\text{UB}(i)} > \Pi_a^{(i)}$}
        \STATE ${\pi}^{\text{O}(i+1)}_{a,g} = {\pi}^{\text{DM}(i)}_{g-1} \: \forall g \in G | {\pi}_{a,g}^\text{True} \leq {\pi}^{\text{DM}(i)}_{g-1}$
    \ELSE
        \STATE $\bm{\pi}^{\textbf{O}(i+1)}_a = \bm{\pi}^{\textbf{O}(i)}_a$
    \ENDIF

  \end{algorithmic}
\end{algorithm}

An underbidding (UB) strategy -- as shown in Algorithm~3 -- implies that an agent underbids its rivals to get more of its offers accepted. In this case the agents need to be informed of the flexibility demand met $\textbf{P}^{\textbf{DM}(i)} = \{\text{P}^{\text{DM}(i)}_g \in \mathbb{R}_{\geq 0} | \text{P}^{\text{DM}(i)}_{g} = \text{P}^{\text{DM}(i)}_{g-1} + \lVert \bm{P}^\textbf{A}_{g} \rVert_1 \}$ at the corresponding price values $\bm{\pi}^{\textbf{DM}(i)} = \{{\pi}^{\text{DM}(i)}_g \in \mathbb{R}_{\geq 0} \}$, where $g \in G$. As with overpricing, agents keep their offer capacity values $\textbf{P}^\textbf{O}_a$ at the maximum, while deciding on the offer prices $\bm{\pi}^\textbf{O}_a$. According to the algorithm, the agents start with the significantly overpriced offers at $\pi^\text{Start}_a$ and decrease them (step 5) if the expected profit from underbidding $\Pi_a^{\text{UB}(i)}$ (found at step 3) is higher than the current one ($\Pi_a^{\text{Ub}(i)} > \Pi_a^{(i)}$). Otherwise, offer prices remain the same (step 7).

\section{Case study}\label{sec:num_study}

This section describes a case study, which is used to demonstrate the proposed two-step modelling approach. 
\textcolor{black}{Fig.~2 illustrates the case study, where a 33/11 kV primary substation that supplies a radial distribution network is experiencing overload in firm capacity. Therefore, to avoid bulk investment in network reinforcement, which would mean a 100\% increase of firm capacity (or 50\% increase of installed capacity), the DSO contracts and dispatches just enough power downstream of the congestion through a demand-side flexibility service to ensure that firm capacity limit is not violated (i.e., performs congestion management). In the particular case study, we consider a real distribution network in the North East of England, where the expected peak demand at the primary substation and the uptakes of flexible assets engaged with the service have been taken from the forecasted distribution future energy scenarios developed and published by local DNO in \cite{dfes2020}.}

\begin{figure}
\centering
        \includegraphics[width=0.6\textwidth]{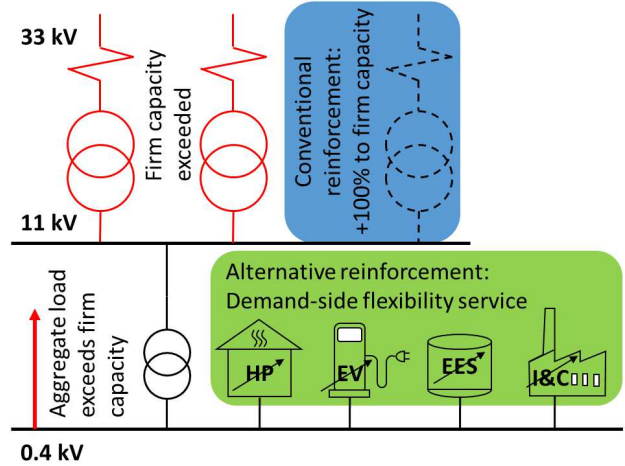}
        \textcolor{black}{\caption{Case study}}
        \label{fig:case_study}
\end{figure}

\textcolor{black}{It is worth noting that the considered network supplies a primarily domestic area, which is already developed, and no additional customers are expected to connect. Hence, the peak demand is directly correlated with the penetration of EVs and HPs. The considered demand-side flexibility service is expected to be delivered by the same assets that cause transformer overload, as well as other flexible assets expected within the network, i.e., EES and I\&C DSR, where the level of engagement determines a particular number of flexible assets that take part in the demand-side flexibility service. The proposed methodology allows us to evaluate whether the considered distribution network can handle the electrification of heating and transport in an economically efficient manner.}

The flexibility service requirements, \textcolor{black}{customer engagement} scenarios (uptake of LCTs and DERs), heating, EV charging demand, EES and I\&C DSR data that correspond to the real network are provided in subsections~\ref{subsec:serv_req}-\ref{subsec:ic_dsr}, respectively. In the numerical study, flexible assets have been distributed between a number of aggregators in a non-uniform way; this was achieved using the function described in subsection~\ref{subsec:flex_dist}.

\subsection{Service requirements}\label{subsec:serv_req}

The flexibility service considered in the case study aims to reduce peak demand at the primary substation to ensure that the net demand does not exceed firm capacity, ensuring a reliable power supply. The network under consideration is a real distribution network \textcolor{black}{of radial topology} in the North East of England, which is supplied by a 33/11 kV primary substation with a firm capacity of 21 MVA. The substation primarily supplies domestic load, which accounts for 15,000 customers. The flexibility service requirements have been derived for this network based on the expected peak demand and peak utilization in the DSO's distribution future energy scenarios \cite{dfes2020} for the year 2030. These requirements are specified in Table~\ref{tab:flex_req}. For the considered flexibility service, the DSO requires a capacity of 2.5 MW to be available every day from 16:30 to 18:30. The maximum price the DSO is willing to pay for the service corresponds to an availability fee of 50 \textsterling/MW/h. This ceiling price represents the cost of conventional network reinforcement.

\begin{table}
\caption{Flexibility service requirements}\label{tab:flex_req}
\begin{center}
\begin{tabular}{l | c | c}
\textbf{Demand capacity} & $\text{P}^\text{D}$ & 2.5 MW\\ 
\textbf{Ceiling price} & $\bar{\pi}$ & 50 \textsterling/MW/h\\ 
\textbf{Flexibility time window} & $T^\text{FW}$ & 16:30 - 18:30\\ 
\end{tabular}
\end{center}
\end{table}

\textcolor{black}{\subsection{Customer engagement scenarios}}

\textcolor{black}{The scenarios of customer engagement with the DSO service} have been taken from distribution \cite{dfes2020} and national \cite{fes2020} future energy scenarios for the year 2030. \textcolor{black}{There are four scenarios of customer engagement considered in the numerical study:} System Transformation (ST), Customer Transformation (CT), Leading the Way (LW), and Net-Zero Early (NZE). \textcolor{black}{The particular values} are \textcolor{black}{given} in Table~\ref{tab:dfes_scen}\textcolor{black}{, where the} scenarios provide the expected numbers of flexible assets (HPs and EVs), installed capacity of EES, and expected capacity of I\&C DSR \textcolor{black}{to be engaged with the DSO contracted flexibility service}. All scenarios are considered to be equally probable.

\begin{table}
\caption{\textcolor{black}{Customer engagement} scenarios for 2030}\label{tab:dfes_scen}
\begin{center}
\begin{tabular}{l | c | c | c | c | c}
\textbf{Type} & Symb                & \textbf{ST}   & \textbf{CT}   & \textbf{LW}   & \textbf{NZE} \\ \hline
EVs, num. & $\text{N}^\text{EV}$    & 2,528         & 5,117         & 6,680         & 7,157  \\ 
HPs, num. & $\text{N}^\text{HP}$    & 782           & 3,454         & 4,207         & 2,015 \\ 
EES, kW & $\bar{\text{P}}^\text{ES}$ & 134         & 236           & 729           & 736 \\ 
I\&C, kW & $\bar{\text{P}}^\text{I\&C}$ & 616       & 901           & 869           & 928 \\ \hline
\end{tabular}
\end{center}
\end{table}

\subsection{Heating demand}
The data required to model flexibility from heating demand is provided in Table~\ref{tab:dwel_type}. It includes parameters for four most common dwelling types in England: Detached (Det.), Semi-Detached (S-Det.), Terraced (Ter.), and Flat. The share of each dwelling type corresponds to the area served by the case study network, while thermal conductance and thermal capacitance for each dwelling type have been derived from the data in \cite{allison2018assessing,hosseinloo2020data}. 

\begin{table}
\caption{Dwelling characteristics}\label{tab:dwel_type}
\begin{center}
\begin{tabular}{l | c | c | c | c | c}
\textbf{Param.} & \textbf{Symb.} & \textbf{Det.} & \textbf{S-Det.} & \textbf{Ter.} & \textbf{Flat} \\ \hline
Index           & $d$            & 1                & 2                   & 3                 & 4 \\ 
Shear, \%       & $\mu_d$        & 6.8              & 34.8                & 30.9              & 27.5  \\ 
Thermal conductance, W/\degree C & $\text{K}^\text{Th}_d$ & 160.3     & 111.4               & 76.4              & 38.1  \\ 
Thermal capacitance, kWh/\degree C & $\text{C}^\text{Th}_d$ & 10       & 6.5                 & 5                 & 4  \\ \hline
\end{tabular}
\end{center}
\end{table}

\subsection{EV charging demand}

The data required to model EV charging demand has been taken from \cite{element2019ev}. It includes the daily average charging demand of 4.8 kWh per EV per day and the average charging station capacity of 6 kW. The share of plugged in EVs during a day has been derived for residential sector from \cite{madzharov2014integrating}.

\subsection{EES data}

The input data for EES modelling has been taken from \cite{sayfutdinov2020alternating}. It includes charge efficiency $\eta^\text{Ch}=97.5\%$, discharge efficiency $\eta^\text{Dis}=97.5\%$, and investment cost for installed capacity $\text{C}^\text{E} = 100$ \textsterling/kWh. The installed energy capacity is assumed to be equal to two hours of discharge at full power. The EES cycle lifetime for different values of DoD is provided in Table~\ref{tab:ees_cyc_lt}.

\begin{table}
\caption{EES cycle lifetime}\label{tab:ees_cyc_lt}
\begin{center}
\begin{tabular}{l | c | c | c | c | c}
DoD     & 10\%   & 20\%   & 30\%   & 40\%  & 50\%  \\ \hline
Cyc.Lt. & 13,660 & 12,200 & 10,800 & 9,480 & 8,230  \\ \hline \hline
DoD     & 60\%   & 70\%   & 80\%   & 90\%  & 100\% \\ \hline
Cyc.Lt. & 7,090  & 6,030  & 5,080  & 4,230 & 3,490 \\ 
\end{tabular}
\end{center}
\end{table}

\subsection{I\&C DSR}\label{subsec:ic_dsr}

Industrial and commercial demand side response (I\&C DSR) is highly dependant on specific business cases and requires detailed modelling of technological processes \cite{papadaskalopoulos2018quantifying}. For example, in \cite{oikonomou2019optimal} Oikonomou and Parvania perform comprehensive modelling of a water desalination plant to determine potential capacity and marginal cost for providing DSR. Since the detailed modelling of businesses and technological processes is out of the scope of the present paper, a quadratic I\&C DSR cost function has been assumed:
\begin{equation}\label{eq:ic_cost}
	C^\text{I\&C}(P^\text{F}) = \text{a}^\text{I\&C} {P^\text{F}}^2 + \text{b}^\text{I\&C} P^\text{F},
\end{equation}
where $\text{a}^\text{I\&C}$ and $\text{b}^\text{I\&C}$ are quadratic and linear terms' coefficients of the cost function that have been derived from the I\&C DSR trial project \cite{clnr2014}. The particular values of I\&C DSR cost function coefficients and other model parameters are provided in Table~\ref{tab:ic_param}.

\begin{table}
	\caption{I\&C DSR model parameters}\label{tab:ic_param}
	\begin{center}
		\begin{tabular}{l | c | c }
			\textbf{Param.} 				& \textbf{Symb.} 			& \textbf{Value} \\ \hline
			Quadratic cost coefficient     	& $\text{a}^\text{I\&C}$ 	& $17.65/\bar{\text{P}}^\text{I\&C}$ * \\
			Linear cost	coefficient			& $\text{b}^\text{I\&C}$ 	& 23.52 \\
			Energy recovery factor			& $\text{e}^\text{I\&C}$		& 1.0 \\ 
			Power recovery factor			& $\text{p}^\text{I\&C}$		& 0.5 \\
			Recovery time period			& $T^\text{Rec}$			& 18:30 - 22:30\\ \hline
		\end{tabular}
	\end{center}
* - $\bar{\text{P}}^\text{I\&C}$ is the maximum capacity of I\&C DSR (Table~\ref{tab:dfes_scen}).
\end{table}

\subsection{Flexible capacity distribution}\label{subsec:flex_dist}
The numerical study considers three types of flexibility provider: domestic, EES, and I\&C. Domestic flexibility providers are those which represent flexibility from HPs and EVs. The other two types of providers represent flexibility from EES and I\&C DSR, respectively. 
The following rule is applied to distribute flexible assets (uptake values of LCTs and DERs) between the agents of the same type:
\begin{equation}\label{eq:share}
    \sigma_{a_n} = \frac{1}{a_n \sum_{j=1}^{|A_n|} \frac{1}{j}},
\end{equation}
where $A_n$ is a set of agents of a particular type $n \in N$ indexed by $a_n$, and $|A_n|$ is its cardinality.

\section{Results}\label{sec:disc}

The results of assessing the case study network using the proposed two-step modelling framework are presented in this section. The results of Step I, where the available flexible capacity and its marginal cost, i.e., offer curves, have been determined for each type of flexible assets and \textcolor{black}{customer engagement} scenario, are presented in section \ref{sec:res1}. The results of Step II, which show the market equilibrium prices for demand-side flexibility service derived using each of the market clearing mechanisms (i.e., pay-as-bid, pay-as-cleared, Dutch reverse auction, and Vickery-Clarke-Groves auction), bidding strategy (i.e., overpricing, capacity understatement and underbidding), and different numbers of agents are presented in section \ref{sec:res2}. Finally, the obtained results have been used for cost-benefit analysis of DSO and flexibility service providers, which is shown in section \ref{sec:res3}.

\subsection{Results of step I - Flexibility supply curves}\label{sec:res1}

\begin{figure}
\centering
        \includegraphics[width=0.75\textwidth]{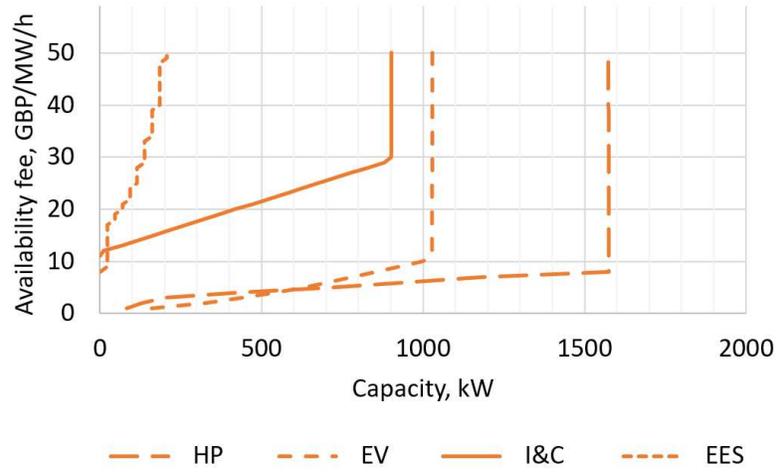}
        \caption{Customer Transformation (CT) scenario: flexibility supply curves}
        \label{fig:CT_flex}
\end{figure}

To derive the available flexible capacity and its marginal cost for each type of flexible assets and \textcolor{black}{customer engagement} scenario, the optimization problems in Section~\ref{sec:flex_models} have been \textcolor{black}{solely} solved for all values of availability fee $\pi = \{1,2,...,\bar{\pi}\}$, where $\bar{\pi}$ is the DSO ceiling price for the flexibility service. The results obtained for the Customer Transformation (CT) scenario are provided in Fig.~\ref{fig:CT_flex}. The figure illustrates the amounts of flexible capacity obtained from the considered flexible assets and their marginal cost, which reflects operation costs for the service delivery. The flexible capacity from any flexible asset for a given availability fee is proportional to the uptake rate of that LCT. Among all LCTs, demand-side flexibility from HPs is the cheapest, being able to provide full potential at only 8 \textsterling/MW/h. The maximum flexibility from EVs can be delivered at 11 \textsterling/MW/h, while for I\&C DSR and EES, the values correspond to 30 and $>$50 \textsterling/MW/h, respectively.

\begin{figure}
\centering
        \includegraphics[width=0.75\textwidth]{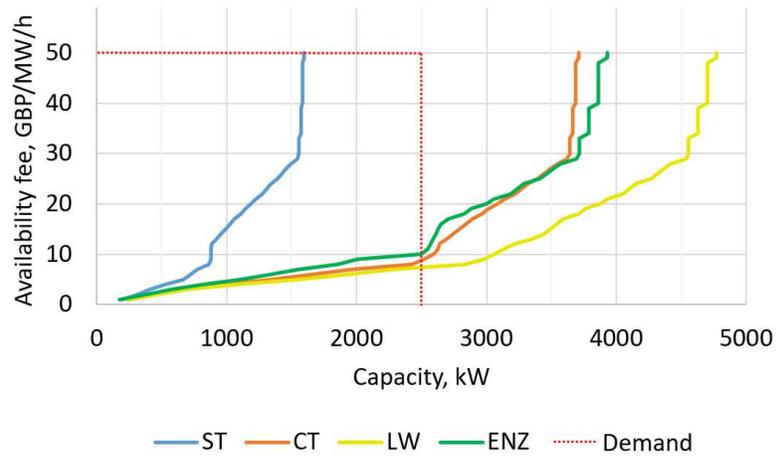}
        \caption{Total flexibility supply curves}
        \label{fig:Tot_flex}
\end{figure}

The aggregate flexibility supply curves for all four \textcolor{black}{customer engagement} scenarios are depicted in Fig.~\ref{fig:Tot_flex}, where the red dotted lines illustrate the DSO demand for flexibility, i.e., required capacity and ceiling price. In three out of the four scenarios, the available flexible capacity is greater than the demand. The intersection of the flexibility demand with supply curves gives the true market equilibrium price for the service. The lowest true market equilibrium price for the flexibility service is observed for Leading the Way (LW) scenario, which equals to 8 \textsterling/MW/h, while the true market equilibrium price for Customer Transformation (CT) and Early Net Zero (ENZ) scenarios correspond to 9 and 10 \textsterling/MW/h, respectively. For the System Transformation (ST) scenario, there is not enough flexible capacity to meet DSO flexibility requirements; hence, the true market equilibrium price equals to the ceiling price (50 \textsterling/MW/h).

\subsection{Results of step II - Strategic bidding and dominant strategies}\label{sec:res2}

\begin{figure*}[h!]
\begin{center}
\subfigure[Pay-as-bid]{\includegraphics[width=0.49\textwidth]{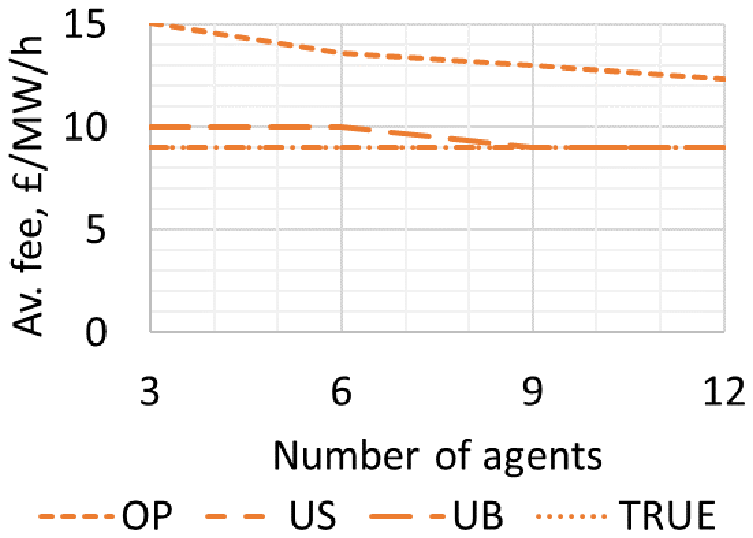}}
\hspace{0.0\textwidth}
\subfigure[Pay-as-cleared]{\includegraphics[width=0.49\textwidth]{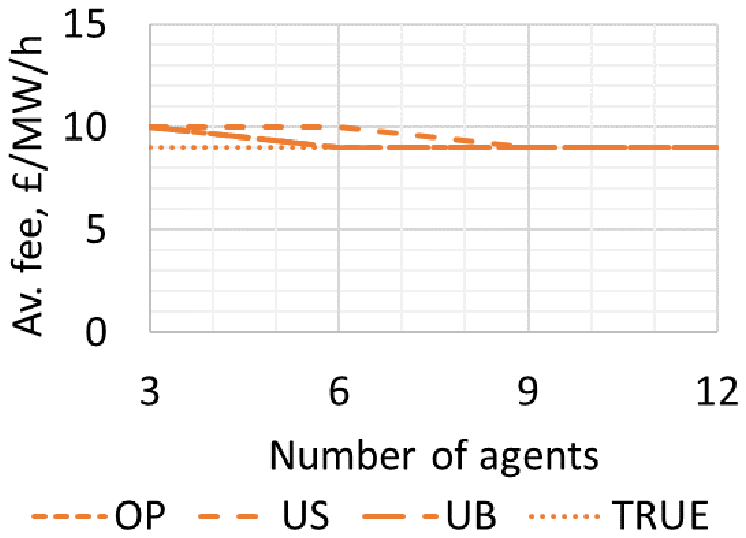}}
\subfigure[Dutch reverse auction]{\includegraphics[width=0.49\textwidth]{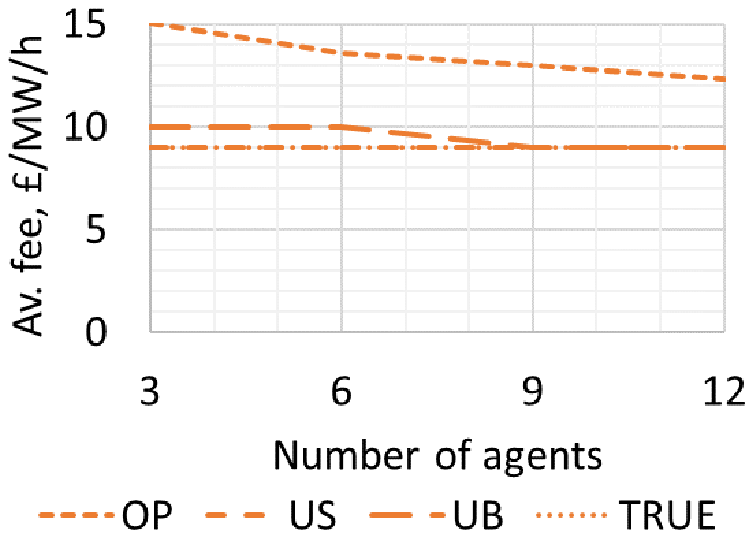}}
\hspace{0.0\textwidth}
\subfigure[Vickrey–Clarke–Groves auction]{\includegraphics[width=0.49\textwidth]{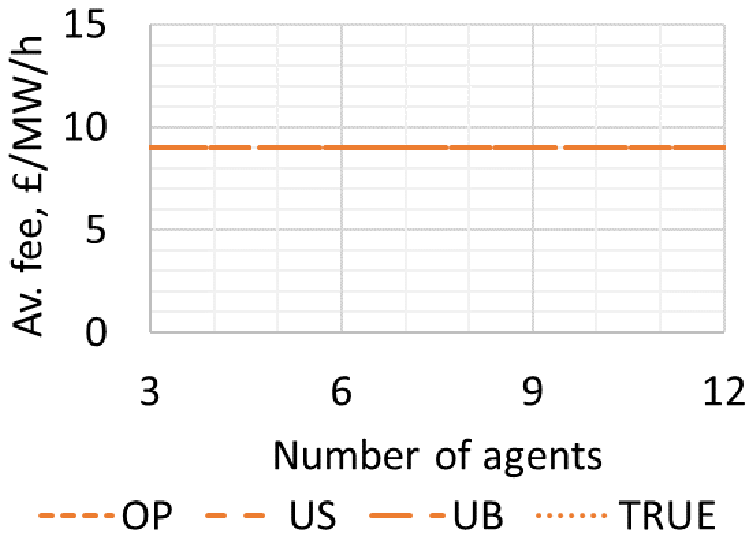}}
\caption{Results of MA modelling for Customer Transformation scenario}
\label{fig:MA_dom}
\end{center}
\end{figure*}

The MA iterative game modelling framework has been applied using the offer curves obtained in section \ref{sec:res1} which correspond to four \textcolor{black}{customer engagement} scenarios. The impact of the number of agents, market clearing mechanism, and bidding strategies are investigated for each scenario. The results for CT electrification scenario, four compositions of agents (3, 6, 9, and 12), three bidding strategies, and four market clearing mechanisms are illustrated in Fig.~\ref{fig:MA_dom}.

Fig.~\ref{fig:MA_dom} shows that each market mechanism has a clear dominant bidding strategy. For example, in the pay-as-bid (PAB) mechanism (Fig.~\ref{fig:MA_dom} (a)), the agents that follow overpricing (OP) strategy are able to maximize the market equilibrium price for the flexibility service compared to the other bidding strategies. For pay-as-cleared (PAC), illustrated in Fig.~\ref{fig:MA_dom} (b), the maximum market equilibrium price is obtained when the agents follow understatement (US) strategy; however, the deviation from the true marginal price is only 1 \textsterling/MW/h and it is only observed for the scenarios with the least competition (three and six agents). For the Dutch reverse auction (DRA), illustrated in Fig.~\ref{fig:MA_dom} (c), the results are identical to the PAB mechanism: both mechanisms are price discriminatory and, under the assumption that all agents follow the same strategy, the result is the same. However, some experimental studies, for example, by Gillen \textit{et al.} \cite{gillen2016bid} showed that due to the specific clearing procedure, the DRA market participants tend to follow the underbidding (UB) strategy, which in this case provides market equilibrium prices which deviate from the true market equilibrium by 1 \textsterling/MW/h for the three and six agents scenarios. The results for the Vickery-Clarke-Groves (VCG) auction, presented in Fig.~\ref{fig:MA_dom} (d), suggest that none of the considered bidding strategies is beneficial for flexibility service providers, which should be expected given the literature. In fact, the objective of MA modelling for the VCG was to establish a benchmark for the cost-benefit analysis provided in section \ref{sec:res3}.

\subsection{Flexibility service cost-benefit analysis}\label{sec:res3}

The expected cost and benefit for both the DSO and flexibility service providers depend on the market clearing (both the equilibrium market price and capacity traded). In this section, only the equilibrium prices that correspond to the dominant strategies, i.e., upper bound of the results illustrated in Fig.~\ref{fig:MA_dom}, have been considered, with the additional study for DRA in which the agents tend to follow the UB strategy, as suggested by \cite{gillen2016bid}. In both discriminatory pricing mechanisms (PAB and DRA) the OP and UB strategies led to the offer prices of the infra-marginal offers approaching market equilibrium price; hence, consideration of market equilibrium prices for the cost-benefit analysis is reasonable for both discriminatory and uniform pricing mechanism.

\begin{figure}
\centering
        \includegraphics[width=0.75\textwidth]{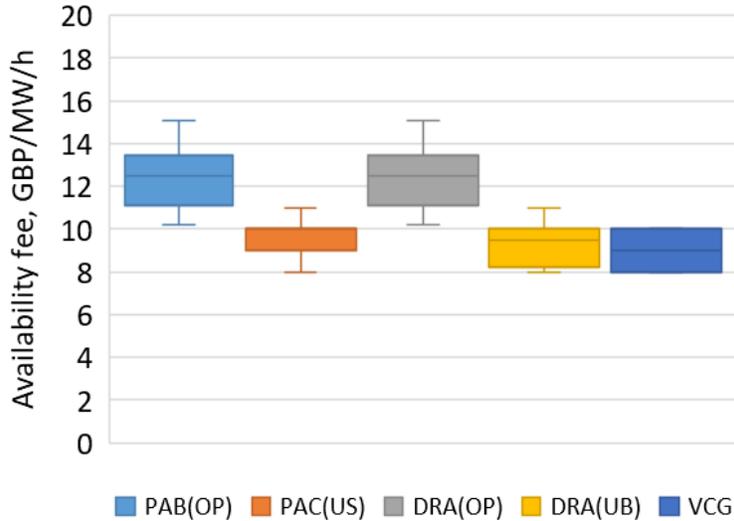}
        \caption{Expected price for flexibility for all electrification scenarios}
        \label{fig:exp_price}
\end{figure}

Box and Whisker plots for the expected equilibrium prices, accounting for strategic bidding, for each market clearing mechanism are depicted in Fig.~\ref{fig:exp_price}. The highest average price (12.4 \textsterling/MW/h) and the highest price variability are expected for PAB and DRA mechanisms, assuming that the flexibility providers follow the overpricing strategy. If the underbidding strategy is chosen for DRA (as was shown in \cite{gillen2016bid}), the expected average price decreases to 9.3 \textsterling/MW/h. The lowest price variability is expected for PAC mechanism under the assumption that the understatement strategy is dominant; the average expected price equals to 9.6 \textsterling/MW/h in this case. The expected true market equilibrium price variation for the flexibility service is provided by the VCG case, for which the average price corresponds to 9 \textsterling/MW/h.

\begin{figure*}
\begin{center}
\subfigure[DSO cost-benefit]{\includegraphics[width=0.475\textwidth]{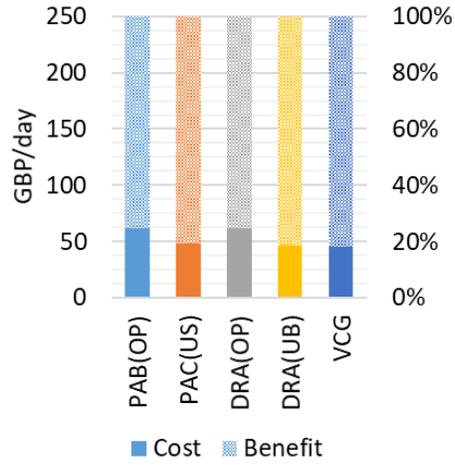}}
\hspace{0.2\textwidth}
\subfigure[Providers' revenue]{\includegraphics[width=0.75\textwidth]{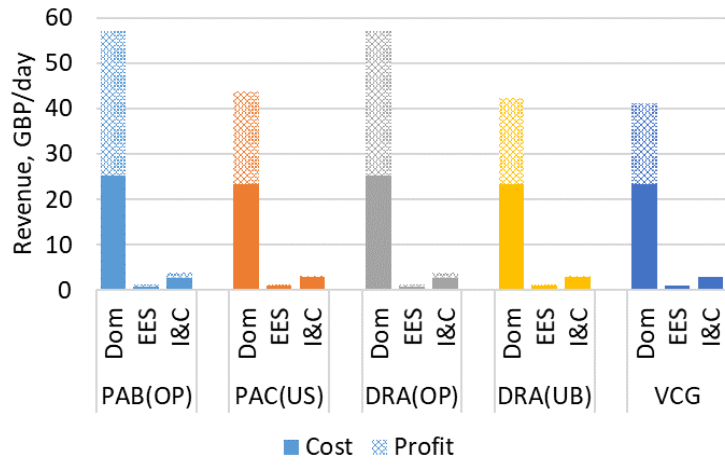}}
\caption{Cost-benefit analysis}
\label{fig:cost_ben}
\end{center}
\end{figure*}

The expected costs and benefits for the DSO, as well as the costs and profits of flexibility providers, have been calculated based on the average expected prices given above. Fig.~\ref{fig:cost_ben} (a) provides the DSO expected costs and benefits for each market clearing mechanism and its dominant bidding strategy. Considering a ceiling price of 50 \textsterling/MW/h, which represents cost for network reinforcement, DSO benefit from the flexibility service varies from 188 to 205 \textsterling/day. The lowest daily benefit corresponds to PAB(OP) and DRA(OP), while the highest corresponds to the strategy-proof VCG, which has not been widely adopted in practice \cite{rothkopf2007thirteen}. The PAC(US) market and DRA(UB) provide 202 and 203.5 \textsterling/day.

Fig.~\ref{fig:cost_ben} (b) depicts the flexibility providers' costs and profits for each market clearing mechanism and bidding strategy. In all cases, domestic flexibility providers have the cheapest resources (HPs and EVs) which deliver the highest revenue, varying from 41.1 to 57 \textsterling/day. The marginal EES and I\&C providers' combined revenue does not exceed 5 \textsterling/day. Looking at the combined providers' profits, the highest profit is observed for the discriminatory pricing mechanisms (PAB and DRA) for OP strategy, which accounts for 53.3\% of the total revenue. When the uniform pricing mechanism is used (PAC(US)), the combined providers' profit makes up 43.3\% of the total revenue. Under the assumption that the agents follow UB strategy in DRA, the providers' profit reaches 41.4\%, while in case of the strategy-proof VCG mechanism, it accounts for 39.5\%

\textcolor{black}{
\section{Discussion}\label{sec:disc}}

\textcolor{black}{
\subsection{Implementation and computational burden}
Fig.~8 illustrates the proposed two-step modelling framework for techno-economic analysis of a demand-side flexibility service, where the two blocks standing on the opposite sides depict input and output data, two blocks in the middle illustrate the two consecutive steps of the proposed modelling approach, and numbered arrows provide the sequence of data feeds. Step I, which is informed by the input data, including heating demand, EV charging demand, EES and I\&C DSR data, customer engagement scenario, and flexibility service time window, individually performs optimization of distributed flexibility models for HP (\eqref{eq:hp_obj}-\eqref{eq:hp_con5}), EV (\eqref{eq:ev_obj}-\eqref{eq:ev_con6}), EES (\eqref{eq:es_obj}-\eqref{eq:es_con9}), and I\&C DSR (\eqref{eq:ic_obj}-\eqref{eq:ic_con3}). For each customer engagement scenario, these optimization problems are solved multiple times for all discrete values of availability fee $\pi=\{1,2,..,\bar{\pi}\}$ to determine the corresponding available flexible capacity $P^\text{F}(\pi)$ from each flexible asset (i.e., flexibility supply curves). The outputs of the first step, along with the flexibility service requirements (i.e., capacity demand and ceiling price) and information about market participants (i.e., number of agents and their shares of flexible capacity), form an input to the second step. Step II performs multi-agent iterative game modelling for combinations of bidding strategies and market clearing mechanisms, where the market clearing problem \eqref{eq:market_obj}-\eqref{eq:market_con2} is solved repeatedly, while agents update their offers following their bidding strategies. The outputs of the second step are equilibrium market clearing prices, and capacities traded, which allow estimating dominant strategies for service providers and performing cost-benefit analysis for DSO and flexibility service providers.}

\begin{figure}
\centering
        \includegraphics[width=1\textwidth]{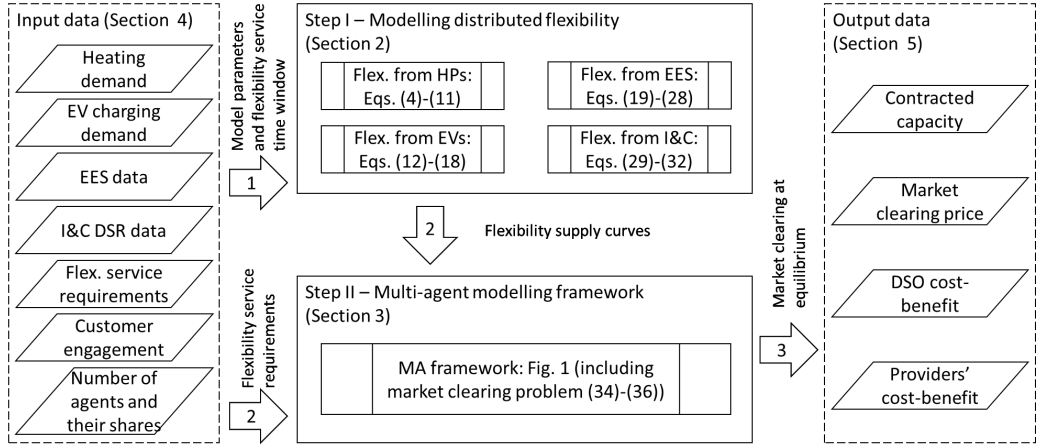}
        \textcolor{black}{\caption{Two-step modelling framework}}
        \label{fig:framework}
\end{figure}

\textcolor{black}{
The proposed modelling framework and individual optimization problems were implemented in JuMP (Julia for Mathematical Optimization). The details of each optimization problem are specified in Table~6, including the type of optimization problem, numbers of real and integer variables, considered time horizon, time resolution, and solver used. For instance, optimization problems for modelling distributed flexibility from HPs, EVs, and I\&C DSR are Convex Programming (CP) problems, while for EES it is mixed-integer linear programming (MILP) problem. The optimization problem for market clearing is of Linear Programming (LP) type. Ipopt solver was used for solving LP and CP problems, while Juniper was used for MILP problem.
}

\begin{table}
	\textcolor{black}{\caption{Specification of optimization problems}}\label{tab:spec_opt}
	\begin{center}
		\begin{tabular}{l | c | c | c | c | c}
\multirow{3}{4em}{Name} & \multicolumn{4}{|c|}{Distributed flexibility models} & \multirow{3}{4em}{Market clearing \eqref{eq:market_obj}-\eqref{eq:market_con2}} \\
& HP & EV & EES & I\&C & \\ 
& \eqref{eq:hp_obj}-\eqref{eq:hp_con5} & \eqref{eq:ev_obj}-\eqref{eq:ev_con6} & \eqref{eq:es_obj}-\eqref{eq:es_con9} & \eqref{eq:ic_obj}-\eqref{eq:ic_con3} & \\ \hline 
Problem type & CP & CP & MILP & CP & LP \\
Real variables & 769 & 193 & 148 & 13 & 1+50$|A|$*\\
Integer variables & - & - & 10 & - & -\\
Time horizon, hour & 24 & 24 & 24 & 6 & -\\
Time step, hour & 0.5 & 0.5 & 0.5 & 0.5 & -\\
Solver used & Ipopt & Ipopt & Juniper & Ipopt & Ipopt \\ \hline
		\end{tabular}
	\end{center}
* - $|A|$ is a total number of agents.
\end{table}

\textcolor{black}{The developed framework has been run on Intel® Core™ i5-7200U CPU @ 2.5GHz 2.71GHz 8GB RAM laptop computer. Each individual optimization problem is solved within a 300 millisecond - 10 second range. The computational time required to solve the case study above was 92 minutes.}

\textcolor{black}{
\subsection{Novel findings}
As it can be seen from the results of the first step, illustrated in Fig.~\ref{fig:CT_flex}, for the case study network, demand-side flexibility that originates from domestic assets (i.e., HPs and EVs) is both significantly higher in terms of capacity and lower in terms of marginal price than EES and I\&C DSR. While the latter solutions have been adopted in the existing networks, domestic flexibility is a new concept that has not been widely adopted in practice. The results illustrate a very high potential for domestic flexibility assets to provide services to local distribution networks, as HPs and EVs can provide significantly higher capacity than the existing solutions (i.e., EES and I\&C DSR) at a lower cost. Ultimately, it is of mutual interest to promote the engagement of domestic customers with the DSO (even through aggregators), as it can provide substantial values to both, which is explicitly illustrated in the cost-benefit analysis depicted in Fig.~\ref{fig:cost_ben}.}

\textcolor{black}{
Regarding the results of the second step, a particular choice of market clearing mechanism for network services is not solely about cost minimization for the DSO. There needs to be a balance between DSO costs and adequate remuneration for service providers, i.e., economic signals. Our study showed that for the PAC case, the combined providers’ profit accounts for 43.3\% of the total revenue, which is considered an adequate incentive. In addition, as it can be seen from Fig.~\ref{fig:exp_price}, the price variation for the PAC is much smaller relative to the other cases, meaning that there is a lower risk of actual service price deviating from the average expected value, giving certainty for DSO planning. Thus, to establish predictable demand-side flexibility service, the DSO should choose PAC mechanism that provides considerable revenue to providers while minimizing possibilities for market manipulations. However, PAB can be chosen initially to provide economic signals and stimulate customer engagement.}

\textcolor{black}{The results of the alternative solutions, illustrated in Fig.~\ref{fig:exp_price}, suggest that with the assumption of underbidding (UB) strategy being dominant for the DRA (previously reported in \cite{gillen2016bid}), the considered demand-side flexibility service would look more beneficial for the DSO in terms of the average equilibrium service price. Specifically, the average equilibrium price for the DRA(UB) case is 9.3 £/MW/h against 9.6 £/MW/h for the PAC(US) case. However, our framework allowed us to identify the vulnerability of the DRA to overpricing (OP) strategy that leads to the significant increase of the average equilibrium price to 12.4 £/MW/h. Otherwise, we would think that, on average, the DRA performs better than the PAC.}

\textcolor{black}{
\subsection{Future work}
In future work, we plan to enhance the proposed multi-agent modelling approach and bidding strategies to accommodate multiple demand-side flexibility services and local markets. This way, we will be able to model future distribution network scenarios where numerous demand-side flexibility services and markets are expected. For instance, in the study reported by the Energy Networks Association (ENA) \cite{ena2018future}, it is anticipated that distribution networks will accommodate local energy markets, DSO and ESO services, as well as other non-SO services (e.g., suppliers’ demand-side flexibility service). Studying simultaneous operation of the above markets and services within a distribution network will be vital in understanding the influence of these services on network operation conditions in the future, which will directly impact customers', the DSO, and whole-system benefits.
}

\section{Conclusions}\label{sec:con}

The paper presented a two-step framework for techno-economic analysis of a demand-side flexibility service in distribution networks built on optimization-based modelling of flexible assets and game-theoretic analysis of a multi-agent system.
A generalized optimization problem formulation has been proposed and applied to various flexible assets (HPs, EVs, EES, and I\&C DSR), to obtain realistic supply curves for demand-side flexibility service. The obtained supply curves have been used as an input to an MA iterative game framework to determine flexibility service market equilibrium prices and capacity traded, while accounting for strategic bidding of participants.
The combined two-step framework accounts for the individual objectives of flexibility providers, technical constraints of flexible assets, customer preferences, market clearing mechanisms, and strategic bidding. Exclusion of any of these factors would result in an incomplete analysis, because they all play a significant role in real life cases.

The framework has been demonstrated using a case study, which considers a real distribution network in the North East of England in which the local DNO is interested in investigating the technical potential and economic effect of a prospective demand-side  flexibility  service, while relying on the time-proven conventional market mechanisms.

The results of the first step suggest that domestic demand-side flexibility from EVs and HPs provide flexible capacity that is both greater and cheaper than that obtained from EES and I\&C DSR. The numerical study showed that in three out of the four \textcolor{black}{customer engagement} scenarios, there was enough flexibility to meet the DSO requirements for secure power supply. \textcolor{black}{Domestic flexibility was found to have a great potential to provide services to local distribution networks; hence, customer engagement needs to be stimulated by the DSO and aggregators.} 

In the second step, MA modelling was carried out using four established market mechanisms: PAB, PAC, DRA, and VCG. The first three mechanisms are potential candidates for a prospective flexibility service, while the fourth is considered as a benchmark for strategy-free market operation. The results suggest that the expected market price for the flexibility service can vary significantly depending on the market clearing mechanism and bidding strategy; while the former can be defined by DSO, the latter is chosen by flexibility providers to maximise their profit. The results of the multi-agent iterative game modelling showed that the discriminatory pricing mechanisms (PAB and DRA) lead to the highest price for the flexibility service if the agents followed an overpricing (OP) strategy. This exceeded the true market equilibrium price by 3-6 \textsterling/MW/h or 33.3\%-66.6\%. Such combinations of market clearing mechanism and bidding strategy allowed providers to achieve profits that exceed their operating costs (53.3\% of the total revenue), far greater than the 39.5\% profit obtained by providers when the strategy proof VCG auction mechanism was used. The pay-as-cleared (PAC) mechanism under the understatement (US) strategy and Dutch reverse auction (DRA) under the underbidding (UB) strategy provided 43.3\% and 41.4\%, respectively. These mechanisms increased the equilibrium price for the service by 1-2 \textsterling/MW/h. In both cases, the true market equilibrium price was obtained when there were at least 9 flexibility providers (agents). \textcolor{black}{Finally, the lowest price variation was observed for the PAC, where with 50\% probability the expected service price lies within 9-10 \textsterling/MW/hour price range.}

The results therefore suggest that to establish a \textcolor{black}{predictable} and attractive mechanism for a flexibility service, a DSO should choose a PAC mechanism that provides considerable revenue to providers, minimizes opportunities for market manipulations (even in the presence of low levels of competition), and avoids the need for network reinforcement, delivering lower energy bills for consumers. \textcolor{black}{However, PAB can be chosen initially to provide economic signals and stimulate customer engagement.}

\section{Acknowledgments}\label{sec:ackn}

This work was conducted under the Customer-Led Distribution System (CLDS) project led by Northern Powergrid (GB) \cite{clds2021}.






\bibliographystyle{elsarticle-num-names}
\bibliography{sample.bib}







\end{document}